\documentclass[10pt,cleanfoot]{asme2ej}

\usepackage{lineno,hyperref,xcolor}
\usepackage[OT2,T1]{fontenc}

\DeclareSymbolFont{cyrletters}{OT2}{wncyr}{m}{n}

\DeclareMathSymbol{\Sha}{\mathalpha}{cyrletters}{"58}

\usepackage{subcaption}
\usepackage{import}
\usepackage{xifthen}
\usepackage{pdfpages}
\usepackage{transparent}
\usepackage{booktabs}
\usepackage[normalem]{ulem}
\usepackage{graphicx}
\usepackage{epsfig}
\usepackage{fancyhdr}
\usepackage{setspace}
\usepackage{helvet}
\usepackage{array}
\usepackage{multicol}
\usepackage{amssymb,amstext,amsmath,bm,cleveref}

\newcommand{\boldface}[1]{\boldsymbol{#1}}  
\newcommand{\bfa}{\boldface{a}}

\newcommand{\bfk}{\boldface{k}}

\newcommand{\bfm}{\boldface{m}}

\newcommand{\bfr}{\boldface{r}}

\newcommand{\bft}{\boldface{t}}

\newcommand{\bfx}{\boldface{x}}

\newcommand{\bfC}{\boldface{C}}

\newcommand{\bfF}{\boldface{F}}

\newcommand{\bfK}{\boldface{K}}
\newcommand{\bfL}{\boldface{L}}

\newcommand{\bfP}{\boldface{P}}

\newcommand{\bfT}{\boldface{T}}

\newcommand{\bfX}{\boldface{X}}

\newcommand{\bfdelta}{\boldsymbol{\delta}}

\newcommand{\bftheta}{\boldsymbol{\theta}}
\newcommand{\bfkappa}{\boldsymbol{\kappa}}
\newcommand{\bflambda}{\boldsymbol{\lambda}}

\newcommand{\bfpsi}{\boldsymbol{\psi}}

\newcommand{\bfvarphi}{\boldsymbol{\varphi}}

\newcommand{\bfnabla}{\boldsymbol{\nabla}}

\newcommand{\calB}{\mathcal{B}}
\newcommand{\calC}{\mathcal{C}}

\newcommand{\calL}{\mathcal{L}}

\newcommand{\calR}{\mathbb{R}}

\newcommand{\dsC}{\mathbb{C}}

\newcommand{\dsZ}{\mathbb{Z}}

\newcommand{\Rset}{\ensuremath{\mathbb{R}}}

\newlength{\boxwidth}
\def\dd{\;\!\mathrm{d}}

\def\btheorem{\begin{theorem}}
\def\etheorem{\end{theorem}}
\def\blemma{\begin{lemma}}
\def\elemma{\end{lemma}}
\def\bproposition{\begin{proposition}}
\def\eproposition{\end{proposition}}
\def\bcorollary{\begin{corollary}}
\def\ecorollary{\end{corollary}}
\def\bdefinition{\begin{definition}}
\def\edefinition{\end{definition}}
\def\bexample{\begin{example}}
\def\eexample{\end{example}}
\def\bremark{\begin{remark}}
\def\eremark{\end{remark}}

\newcommand{\be}{\begin{equation}}
\newcommand{\ee}{\end{equation}}
\newcommand{\beq}{\begin{eqnarray}}
\newcommand{\eeq}{\end{eqnarray}}
\newcommand{\bem}{\begin{multline}}
\newcommand{\eem}{\end{multline}}
\newcommand{\ba}{\begin{align}}
\newcommand{\ea}{\end{align}}

\renewcommand{\figurename}{Figure}
\renewcommand{\tablename}{Table}

\usepackage{hyperref} 
\hypersetup{
	colorlinks=true,
	linkcolor=blue,
	citecolor=blue,
	urlcolor=blue,
}
\usepackage[square,numbers]{natbib}

\title{Topology optimization of graded truss lattices based on on-the-fly homogenization}

\author{Bastian Telgen
    \affiliation{	Mechanics \& Materials Lab \\ 
    				Department of Mechanical and Process Engineering\\ 
    				ETH Z\"urich \\ 
    				8092 Z\"urich, Switzerland \\ 
    				Email: telgenb@ethz.ch}	
}

\author{Ole Sigmund  
	\affiliation{	TopOpt group \\ 
					Department of Mechanical Engineering \\ 
					Technical University of Denmark \\ 
					2800 Kongens Lyngby, Denmark \\ 
					Email: sigmund@mek.dtu.dk    }
}

\author{Dennis M. Kochmann\thanks{Corresponding author.}
    \affiliation{	Mechanics \& Materials Lab \\ 
    				Department of Mechanical and Process Engineering\\ 
    				ETH Z\"urich \\ 
    				8092 Z\"urich, Switzerland \\ 
    				Email: dmk@ethz.ch}
}

\begin{document}

\maketitle    
\doublespacing

\begin{abstract}
We introduce a computational framework for the topology optimization of cellular structures with spatially varying architecture, which is applied to functionally graded truss lattices under quasistatic loading. We make use of a first-order homogenization approach, which replaces the discrete truss by an effective continuum description to be treated by finite elements in a macroscale boundary value problem. By defining the local truss architecture through a set of Bravais vectors, we formulate the optimization problem with regards to the spatially varying basis vectors and demonstrate its feasibility and performance through a series of benchmark problems in 2D (though the method is sufficiently general to also apply in 3D, as discussed). Both the displacement field and the topology are continuously varying unknown fields on the macroscale, and a regularization is included for well-posedness. We argue that prior solutions obtained from aligning trusses along the directions of principal stresses are included as a special case. The outlined approach results in heterogeneous truss architectures with a smoothly varying unit cell, enabling easy fabrication with a tunable length scale (the latter avoiding the ill-posedness stemming from classical nonconvex methods without an intrinsic length scale).
\end{abstract}

\section{Introduction}
\label{sec:introduction}
Cellular materials achieve excellent mechanical properties at a considerably lower weight than conventional solids. Having been characterized extensively with regards to the wide material property space they span \citep{gibson_ashby_1997,ashby_materials_2016}, cellular metamaterials enable their functional optimization through a careful choice of the base material(s) and, moreover, the topology and geometry of the cellular design \citep{evans_topological_2001}. The latter particularly applies in hierarchical networks, whose overall properties depend on the arrangement of structural members at each level \citep{lakes_materials_1993}. The rapid development of modern additive manufacturing techniques has enabled the integration of cellular design principles across length scales down to the nanoscale \citep{zheng_ultralight_2014,meza_resilient_2015}, which has opened an unprecedented design space. 
Cellular solids are also omnipresent in nature (e.g., in bone, wood, shell, or bamboo), whose key distinguishing feature from most man-made metamaterials is their non-periodicity stemming from self-assembly processes. The design of metamaterials, by constrast, typically explores periodic architectures in engineering design, owing to the simplicity of extracting their effective properties by homogenization.

Truss lattices are one such subclass of periodic cellular materials, exploited extensively over the past few decades  \citep{fleck_micro-architectured_2010,schaedler_architected_2016, valdevit_introduction_2018}. A multitude of architected truss \mbox{(meta-)}materials has been introduced theoretically and experimentally with as-designed properties. 
Classical examples include extreme properties such as designs with high-strength- and/or stiffness-to-weight ratios \citep{schaedler_ultralight_2011}, further unconventional properties such as auxeticity \citep{lakes_foam_1987, sigmund_materials_1994} or near-infinite bulk-to-shear modulus ratios \citep{sigmund_materials_1994, MiltonCherkaev1995,huang_topological_2011,BueckmannEtAl2014}. Advancing from linear to nonlinear and dynamic material behavior, properties of interest have included energy absorption \citep{wadley_compressive_2008}, acoustic wave tuning \citep{HusseinEtAl2014,ZelhoferKochmann2017}, controllable nonlinear stress-strain behavior and shape recoverability under large deformations \citep{clausen_topology_2015, meza_resilient_2015}, insensitivity to imperfections \citep{symons_imperfection_2008, montemayor_insensitivity_2016, thomsen_buckling_2018}, and damage tolerance \citep{pham_damage-tolerant_2019}.

The almost infinite freedom in engineering the topology of cellular materials and the resulting enormous potential reflected by the above examples highlights the challenging inverse problem. Required are systematic approaches to obtain structures with desired properties. When it comes to optimizing the effective material behavior of periodic networks, a multitude of analytical \citep{ evans_topological_2001}, numerical \citep{sigmund_materials_1994,cadman_design_2013, osanov_topology_2016} and empirical methods \citep{jain_commentary:_2013} have been deployed, successfully reaching, e.g., theoretical limits such as stiffness upper bounds \citep{sigmund_new_2000,BergerEtAl2017} 
or demonstrating extreme band gaps in their dispersion relations \citep{sigmund_systematic_2003, bilal_ultrawide_2011}. The integration of such periodic architectures into real-world applications (from lightweight space structures to efficient infills or patient-specific medical devices) is another challenge, since complex macroscale load conditions and constraints must be accounted for \citep{valdevit_introduction_2018,pasini_guest_2019,kochmann_hopkins_valdevit_2019}, so uniform periodic structures are not a solution, and structural-level topology optimization becomes a method of choice.

While topology optimization has evolved into a mature systematic design approach for structural optimization, the classical approach faces high computational costs in case of large-scale optimization problems involving several length scales. The recent benchmark presented in \cite{aage_giga-voxel_2017} proved the impressive capability of a modern high-performance implementation to optimize large structures in detail (based on more than one billion finite elements). If unregularized, the optimization problem lacks existence of solutions, meaning that increasing resolution leads to the appearance of finer and finer features (for a detailed summary on regularization approaches and manufacturability constraints, the reader is referred to \cite{LazWanSig16}). In fact, for linear elastic problems with isotropic phases and multiple loads, the optimal local solution is given by laminations at multiple length-scales (3 in 2D and 6 in 3D) \citep{milton_modelling_1986}. However, in a recent 2D microstructure optimization study \citep{traff_simple_2018} it was demonstrated that single-scale realizations can be obtained at low loss in performance. So-called homogenization-based approaches that use knowledge of optimal microstructures offer great potential for time savings. Such procedures require a separation of scales between the macroscale boundary value problem and the fine-scale microstructure. However, experience \citep{groen_homogenization-based_2018} shows that the optimal local material orientations resulting from such approaches ensure stretching-dominated microstructures that are less prone to scale separation issues.

A number of hierarchical multiscale topology optimization schemes have been introduced, which optimize a macroscale problem while accounting for the microscale topology by homogenization \citep{rodrigues_hierarchical_2002, coelho_hierarchical_2008, sivapuram_simultaneous_2016, wang_multiscale_2017} or computational FE$^2$ techniques \citep{xia_concurrent_2014}. Here, the optimization problem on the macroscale depends on a parametrization of the structural architecture on the lower scale, and the latter is optimized in a spatially varying fashion. Such schemes have often led to locally optimized (and spatially varying) unit cells (e.g., at the quadrature-point level) without ensuring proper compatibility and connectivity across the macroscale body, thus also lacking manufacturability. Different solutions have been proposed to assure the connectivity of spatially varying lattice topologies. In \cite{khanoki_multiscale_2012} and \cite{wang_hip_2018} a square-/tetrahedron-based topology description linked to a constant quadrilaterial/tetrahedral mesh of the design domain in 2D and  was used, respectively, with varying unit cell densities. An appealing generalization to allow for microstructural topology changes is the introduction of microstructural families, in which all members are interconnectable \citep{schumacher_microstructures_2015, chen_computational_2018}. In \cite{radman_topology_2013} non-designable connector cells to link adjacent cells were defined. In \cite{zhu_two-scale_2017} and \cite{du_connecting_2018} a connectivity index used as a constraint during optimization was introduced. In \cite{panetta_elastic_2015} different microstructures were stitched to have a better match, and in \cite{schury_efficient_2012}  transition layers that are optimized for optimal connectivity were introduced. All of these approaches produce elegant structures; however, the fixes come at the expense of a restricted design space and, more severely, the way different unit cells are connected may violate the homogenization assumptions of scale separation and periodicity. 

An elegant alternative is the introduction of microstructures that are continuously graded in space -- analogous to the transition from classical multi-material topology optimization defining a unique material at each point (e.g., \cite{sanders_multi-material_2018, zhang_adaptive_2020, sanders_optimal_2021}) to a continuous grading of materials  \citep{dunning_simultaneous_2015}.
In \cite{greifenstein_simultaneous_2016} a framework for free material optimization with constrained grading was presented and demonstrated its performance for rank-2 laminates; unfortunately, this approach lacked a projection onto a finite length scale. Recently the approaches  shown in \cite{groen_homogenization-based_2018}, \cite{allaire_topology_2019}, \cite{xu_topology_2021} and \cite{kumar_density-and-strain-based_2020}, in line with earlier work in \cite{pantz_post-treatment_2008}, were used to construct spatially variant lattices on a finite length scale, based on a rectangular microstructure representation for which a homogenization-based surrogate model was used. A Similar hybrid approach has also been applied in 3D as presented in \cite{wu_design_2019}. A recent review of such multi-scale topology optimization approaches can be found in \citep{wu_topology_2021}.

In summary, identifying optimal structural architectures with fine-scale features for complex macroscale boundary value problems can be tackled by either high-performance topology optimization schemes on the macroscale (but those come with significant expenses) or by two-scale optimization approaches (which often lack compatibility or a highly limited design space). 
We here present a computational framework which aims at establishing compatible, manufacturable truss networks with a tailored length scale suitable for complex macroscale problems. Based on a chosen microstructure parametrization, we employ a semi-analytical homogenized continuum model for truss lattices accounting for finite rotations \citep{glaesener_continuum_2019}, which is used within a macroscale finite element setup (using both translations and rotations as primary fields). Based on the assumption of a continuously varying truss architecture on the macroscale, our framework allows for the optimization of spatially variant cellular lattice structures.
A projection of the obtained set of truss unit cells at the macroscale onto a chosen finite length scale \citep{rumpf_synthesis_2012} results in the final truss design, thereby assuring manufacturability and respecting the assumptions of periodicity and scale separation. The general framework is applicable to, in principle, any periodic cellular material and hence enables the structural realization, combination and symbiosis of a vast range of microstructural topologies within a single body. 
We demonstrate the approach by seeking functionally-graded truss architectures having minimal compliance. Inspired by the atomic arrangement in crystals, we parametrize the truss unit cell via a set of Bravais lattice vectors along with Voronoi tessellation.
Overall, our approach may be viewed as a generalization of those introduced previously in \cite{groen_homogenization-based_2018}, \cite{GroStuSig19}, and \cite{allaire_topology_2019}.  Other recent approaches in the field include those in \citep{Xue2020} and \citep{GeoAllOli20}.

In the following \Cref{sec:methodology}, we summarize the optimization methodology in four steps. First, \Cref{subsec:spatiallyvariantlattices} introduces the microstructure parametrizaion, followed by a brief summary of the homogenization approach in \Cref{subsec:macromodel} and of the topology optimization problem in \Cref{subsec:topologyoptimzation}. Lastly, the projection onto the finite length scale is presented in \Cref{subsec:continuoustodiscrete}. To demonstrate the performance of this framework, \Cref{sec:numerical_examples} presents a number of benchmarks for functionally graded truss lattices with minimized compliance. Finally, \Cref{sec:conclusion} concludes this work and discusses its benefits and limitations.

\section{Methodology}
\label{sec:methodology}

\begin{figure}[ht]
    \centering
    \includegraphics[width=.8\linewidth]{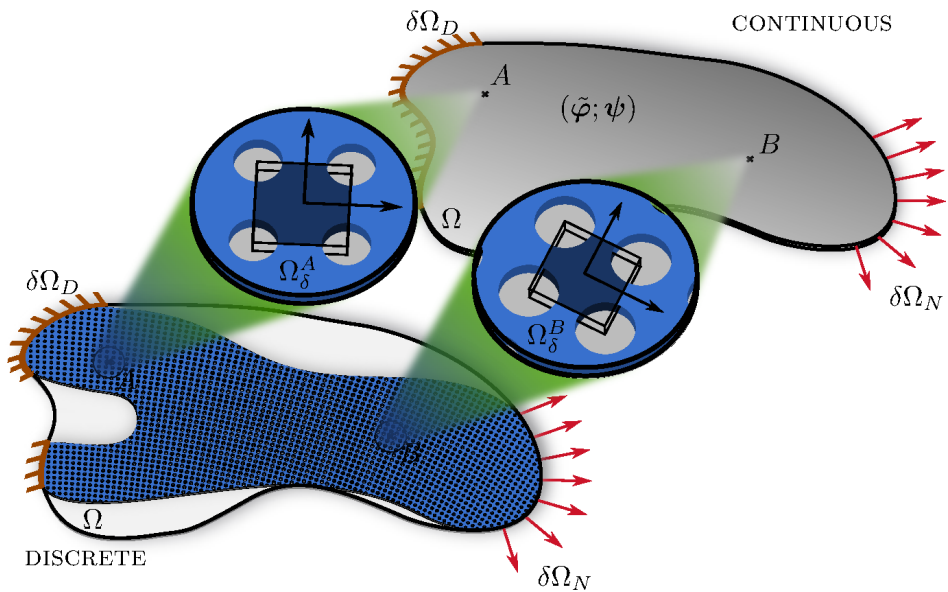}
    \caption{The \textit{discrete} spatially variant structure is described on the macroscale as a mechanically equivalent \textit{continuous} body, whose local effective constitutive behavior derives from the local definition of the truss topology, parameterized by the set $\boldsymbol{\psi}$ of all spatially varying topology variables and based on the assumption of a (quasi-)periodic lattice. The continuously varying topology variables $\boldsymbol{\psi}$ are optimized over the macro-domain $\Omega$ towards a minimum compliance for given boundary conditions (considering Neumann and Dirichlet boundaries $\delta\Omega_N$ and $\delta\Omega_D$, respectively). The final discrete truss representation is obtained by introducing a mapping $\tau'$, which projects the continuous topology variable field $\boldsymbol{\psi}$ onto a discrete structure of finite length scale.}
    \label{fig:method}
\end{figure}

The present optimization methodology is based on a two-scale description of the cellular body, which separates the macroscale boundary value problem from the underlying discrete structural architecture. By assuming a separation of scales, the small-scale truss network is condensed into an effective, homogenized continuum model to be used at the larger scale. The connecting link between both is the definition of the truss topology, which is defined on the macroscale in a spatially varying fashion. Like the displacement field as the primary unknown of the macroscale boundary value problem, the truss topology is continuously varied, and the local unit cell at any point on the macroscale is the basis for the homogenized mechanical response at that point. \autoref{fig:method} schematically illustrates the two-scale setup to be explained in the following.

\subsection{Spatially variant truss topology parametrization}
\label{subsec:spatiallyvariantlattices}

\begin{figure}
	\centering
	\begin{minipage}[t]{0.48\textwidth}	
		\centering
        \includegraphics[width=.75\linewidth]{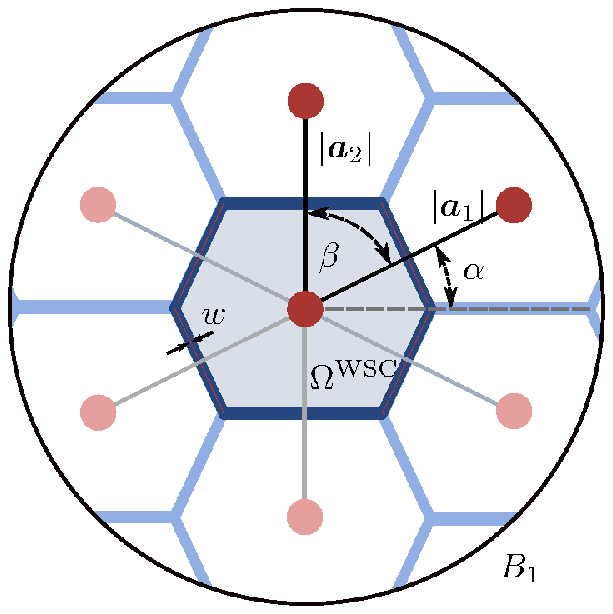}
		\caption{Construction of a unit-volume RUC in an $\varepsilon$-neighborhood $B_1$ around point $\bfX\in\Omega$. Based on the Bravais lattice parameters $\alpha$, $\beta$, and $\gamma = |\bfa_2|/|\bfa_1|$ the lattice sites (red), and hence the Wigner-Seitz cell $\Omega_\text{WSC}$, are uniquely defined. We choose to place nodes $\calB_n^\text{RUC}$ and beams $\calB_b^\text{RUC}$ (blue) on the RUC's corners and edges, respectively. All beam properties are collected in the set $\bflambda$.}
		\label{fig:RepresentativeUnitCell}
	\end{minipage}%
	\hfill
	\begin{minipage}[t]{0.48\textwidth}
		\centering
        \includegraphics[width=.95\linewidth]{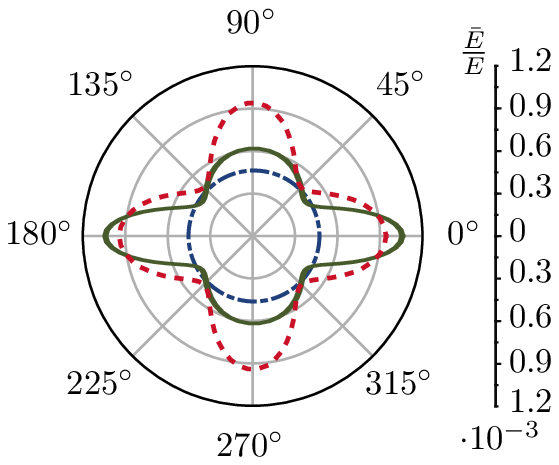}
		\caption{Effective directional Young's modulus vs. orientation for three different lattices at $\rho^{RUC}=0.1$: the tetragonal lattice with a square RUC (dotted red, $\alpha=0$, $\beta=\pi/2$, $\gamma=1$), the orthorhombic lattice with a rectangular RUC (solid green, $\alpha=0$, $\beta=\pi/2$, $\gamma=2$), and the hexagonal lattice with a hexagonal RUC (dash-dotted blue, $\alpha=0$, $\beta=\pi/3$, $\gamma=1$). }
		\label{fig:ElasticSurfaces}
	\end{minipage}
\end{figure}

For the geometric description of spatially variant truss lattices, we borrow principles from (spatially invariant) crystal structures in solid state physics \citep{kittel_introduction_1986}.
Confining ourselves to simple Bravais lattices \cite{bravais_memoire_1850}, each site in an infinite lattice in $d$ dimensions is located at position
\begin{equation}
    \bfr^\prime = \bfr + \sum_{i=1}^d n_i \bfa_i  \qquad \textrm{with} \qquad n_i \in \dsZ,
    \label{eq:BravaisLattice}
\end{equation}
where $\bfr$ denotes a reference point, $\{\bfa_1,\ldots,\bfa_d\}$ are the Bravais basis of linearly independent unit vectors, and $\{n_1,\ldots,n_d\}$ is a unique set of integers. 
Towards the description of spatially variant structures, we now allow the primitive translation vectors to vary smoothly in space on the macroscale, introducing the (at least $\calC^1$-continuous) primitive translation vector fields $\bfa_i(\bfX)$ ($i=1,\ldots,d$), where $\bfX\in\Omega$ denotes a position on the macroscale body $\Omega$ in the undeformed configuration. In an $\varepsilon$-neighborhood of a point $\bfX\in\Omega$, the local lattice site locations are now approximated by
\begin{equation}
    \bfr^\prime \approx \bfX + \sum_{i=1}^d n_i \bfa_i(\bfX)  \qquad \textrm{with} \qquad n_i \in B_\varepsilon[0] \subseteq \dsZ,
    \label{eq:SpatiallyVariantBravaisLattice}
\end{equation}
where $B_\varepsilon[x]  = \{ y\in\dsZ: |x-y| \leq \varepsilon \}$ defines a closed $d$-ball of integer radius $\varepsilon>0$. Since the approximation \eqref{eq:SpatiallyVariantBravaisLattice} is only suitable if the basis vectors and hence the lattice spacing must be significantly smaller than any macroscopic fields of interest ($|\bfa_i|\ll|\Omega|$), we require a separation of scales between the fields describing the structural topology (here, $\bfa_i(\bfX)$) and all macroscopic mechanical fields of interest (such as the displacement field). The topology variations are consequently so smooth on the macroscale that at each point $P\in\Omega$ the underlying structure within an $\varepsilon$-neighborhood can be approximated by a periodic lattice. Naturally, a spatially constant structure (i.e., a perfect lattice) is recovered on the macrscoale if $\bfa_i(\bfX)=\text{const}.$ for all $\bfX\in\Omega$, in which case \eqref{eq:SpatiallyVariantBravaisLattice} is exact for $\varepsilon\to\infty$ and coincides with \eqref{eq:BravaisLattice}. 

The above Bravais construction only defines a point set. In order to arrive at a truss topology, one could potentially connect those sites by beams -- this approach, however, is non-unique and rather restrictive in parametrizing the structural topology. 
Therefore, we instead use the Wigner-Seitz construction\footnote{We point out that this choice is only one possible option among various others (e.g., one could extend the unit cell definition to include several particles, as originally done for multi-lattices in \cite{kittel_introduction_1986}). The general approach laid out here is independent of the specific choice of topology parametrization.} \citep{wigner_constitution_1933} -- based on the set of lattice sites -- to define a unique primitive unit cell topology for a given set of Bravais basis vectors $\{\bfa_1,\ldots,\bfa_d\}$. The Wigner-Seitz cell $\Omega_{\textrm{WSC}}$ is defined as the locus of points in space closer to one lattice site than to any other lattice site (see \autoref{fig:RepresentativeUnitCell}). Its geometric construction can conveniently and efficiently be carried out by Voronoi tessellation over the neighborhood of lattice sites defined in \eqref{eq:SpatiallyVariantBravaisLattice} around any point $\bfX\in\Omega$. The Wigner-Seitz cell serves as representative unit cell (RUC) $\Omega_{\textrm{RUC}}=\Omega_{\textrm{WSC}}$ (see \autoref{fig:RepresentativeUnitCell}), whose $n_n^\textrm{RUC}$ outer vertices and $n_b^\textrm{RUC}$ edges are identified as nodes and beams of a truss, respectively (and we denote the sets of all nodes and of all beams in the RUC by $\calB_n^\textrm{RUC}$ and $\calB_b^\textrm{RUC}$, respectively). This completes the unique definition of the truss unit cell topology based on a given set of (local) Bravais basis vectors. 

\sloppy For convenience, we represent the Bravais lattice vectors in fractional coordinates, i.e., in 2D $\bfa_1=(|\bfa_1| \cos\alpha, |\bfa_1| \sin\alpha)$ and  $\bfa_2=(|\bfa_2| \cos(\alpha+\beta), |\bfa_2| \sin(\alpha+\beta))$, where angles $\alpha$ and $\beta$ define, respectively, the orientation and the relative orientation of the two Bravais vectors (see \autoref{fig:RepresentativeUnitCell}). The scale separation allows us to define the homogenization problem on any scale. To ensure RUC uniqueness, we enforce the representative unit cell to have a normalized volume of $|\Omega_\text{RUC}|=|\text{det}(\bfa_1,\ldots,\bfa_d)|=1$. The volume constraint reduces the number of independent Bravais lattice variables to $d^2-1$. In 2D, we choose $\alpha$, $\beta$, and $\gamma = |\bfa_2|/|\bfa_1|$.

Since the mechanical response of a truss depends not only on topology but also on its material parameters and strut geometry, we here assume that all beams are of rectangular cross-section with constant out-of-plane thickness $t$ and spatially varying width $w(\bfX)$. (It is straightforward to relax this constrained by introducing additional geometric design variables.) The beam width $w$ is a direct consequence of the effective density $\bar{\rho}(\bfX)$ through $w=\bar{\rho}(\bfX)|\Omega_\text{RUC}| / L$. All beams are assumed to be composed of the same isotropic linear elastic material with Young's modulus $E$ and mass density $\rho$. The beam properties are summarized in the set $\bflambda(\bfX)=\{E,\rho,t,w(\bfX)\}$, which in the spatially variant context varies with position $\bfX\in\Omega$ on the macroscale, analogous to the basis vectors. 
From now on, we refer to the set of all design variables, uniquely defining the associated RUC $\Omega_\text{RUC}$ as well as $\calB_n^\textrm{RUC}$ and $\calB_b^\textrm{RUC}$, as
\begin{equation}\label{eq:designparams}
    \boldsymbol{\psi}(\bfX)=\{\bfa_1(\bfX),\ldots,\bfa_d(\bfX);\bflambda(\bfX)\}.
\end{equation}
Any further quantities required for the subsequent optimization problem, such as the local mass density $\rho(\bfX)$ or the effective mechanical response (of an equivalent periodic structure), can be extracted from a given set $\boldsymbol{\psi}(\bfX)$ via homogenization, as described in the following.

\subsection{Effective continuum description based on first-order homogenization}
\label{subsec:macromodel}

The topology optimization problem requires information about the local mechanical constitutive behavior on the macroscale, which must depend on the truss architecture on the microscale. Our natural solution is a first-order homogenization approach, exploiting the assumed separation of scales and extracting the effective response from the RUC.
We here employ the semi-analytical approach in \cite{glaesener_continuum_2019}, whose framework provides an efficient description of periodic truss lattices of general linear elastic beam architectures, accounting for finite rotations and hence geometric nonlinearity. In the following, we summarize the homogenization technique to the extent necessary for subsequent discussions and discuss its validity for the investigated type of spatially variant trusses. For further details of the model and its numerical implementation we refer to \cite{glaesener_continuum_2019}.

Acknowledging that the deformation of Timoshenko-Ehrenfest beams is described by both rotations and translations, we employ a macroscale kinematic continuum description of micropolar type, introducing two primary fields on the macroscale: a deformation mapping $\bfx=\bfvarphi(\bfX)$ and a rotation field $\bftheta=\bftheta(\bfX)$, where $\bfx\in\varphi(\Omega)$ and $\bfX\in\Omega$ are the deformed and undeformed positions of a material point. Assuming the deformation fields vary smoothly (at least $\calC^1$-continuous), the local deformation within an infinitesimal neighborhood of a point $\bfX\in\Omega$ is approximated to first order\footnote{For reduced computational complexity, we choose to truncate the Taylor series to linear order. A detailed discussion and a comparison with second-order approximations can be found in \cite{glaesener_continuum_2019}.} by
\begin{equation}\label{eq:Taylor}
    \begin{split}
        x_i (\bfX+\dd\bfX) = \varphi_i(\bfX) + F_{iJ}(\bfX) \,\dd X_J + \text{h.o.t.},\\
        \theta_i (\bfX+\dd\bfX) = \theta_i(\bfX) + \kappa_{iJ}(\bfX) \dd X_J + \text{h.o.t.},
    \end{split}
\end{equation}
with deformation gradient $\bfF(\bfX)=\nabla \bfvarphi (\bfX)$ and curvature tensor $\bfkappa(\bfX)=\nabla \bftheta (\bfX)$. Under the assumption of a separation of scales, we apply \eqref{eq:Taylor} in the neighborhood of $\bfX\in\Omega$ to the nodes in the truss RUC centered at $\bfX$ and defined by $\boldsymbol{\psi}(\bfX)$. Consequently, every node $\alpha\in\calB_n^\textrm{RUC}$ initially located at $\bfX_\alpha^m\in\Omega_{\textrm{RUC}}$ is deformed according to
\begin{equation}\label{eq:TaylorExpansionMulti}
    \begin{split}
        \bfx_\alpha^m = \bfvarphi(\bfX) + \bfF(\bfX) \bfX_\alpha^m + \delta\bfx_{\alpha} + \text{h.o.t.},\\
        \bftheta_\alpha^m = \bftheta(\bfX) + \bfkappa(\bfX) \bfX_\alpha^m + \delta\bftheta_{\alpha} + \text{h.o.t.}.
    \end{split}
\end{equation}
Here and in the following, the superscript $^m$ denotes microscale quantities. Importantly, unlike the affine deformation in~\eqref{eq:Taylor}, the approximations \eqref{eq:TaylorExpansionMulti} introduce shifts $\tilde\bfdelta_\alpha=(\delta\bfx_{\alpha}, \delta\bftheta_{\alpha})$, which allow nodes to deform relative to the RUC average in a periodic fashion. By defining one set of shifts for each unique periodic node in the unit cell, this effectively imposes periodic boundary conditions for both nodal translations and rotations and hence ensures that the average primary fields ($\bfvarphi$ and $\bftheta$) and their gradients ($\bfF$ and $\bfkappa$) are imposed as averages across the RUC. The relaxation from an affine to a general periodic deformation of the RUC proves to be especially important for bending-dominated truss architectures, in which soft modes are accommodated primarily by bending in a manner far from affine deformation~\citep{glaesener_continuum_2019}. 

Based on the co-rotational beam description in \cite{crisfield_consistent_1990}, the strain energy of each beam in the RUC is a function of the nodal displacements and rotations, such that an effective strain energy density can be defined as the RUC average
\begin{equation}
    W(\boldsymbol{\psi};\bfF;\bftheta,\bfkappa;\tilde{\bfdelta}) =
    \frac{1}{V_\textrm{RUC}} \sum_{\nu \in \calB^\textrm{RUC}_b(\boldsymbol{\psi}) } W_\nu(\tilde{\bfx}_{\nu_1}^m, \tilde{\bfx}_{\nu_2}^m),
\end{equation}
where $V_\textrm{RUC}=|\Omega_\textrm{RUC}|$ is the volume (or area in 2D) of the RUC, and $W_v$ is the strain energy of beam $\nu$ having the two nodes $\nu_1$ and $\nu_2$ with degrees of freedom $\tilde{\bfx}_{\alpha}=(\bfx^m_\alpha,\bftheta^m_\alpha)$ from~\eqref{eq:TaylorExpansionMulti} for $\alpha=\nu_1,\nu_2$. We note that the resulting energy density $W$ is effectively independent of $\bfvarphi$, since $\bfvarphi(\bfX)$ in~\eqref{eq:TaylorExpansionMulti} merely constitutes rigid-body translations, which do not affect the strain energy (whereas $\bftheta(\bfX)$ implies nodal rotations and beam bending, hence not implying rigid-body rotations).

For linear elasticity, the shifts are expected to minimize the unit cell energy (which is local in the shifts), so that we may condense them out on the microscale. The resulting condensed energy density may be stated as
\begin{equation}\label{eq:Wstar}
    W^*(\boldsymbol{\psi};\bfF;\bftheta,\bfkappa) = \min_{\tilde{\bfdelta}} \big\{ W(\boldsymbol{\psi};\bfF;\bftheta,\bfkappa;\tilde{\bfdelta}) \ : \ \text{constr.}\big\},
\end{equation}
with constraints ensuring that the shifts $\tilde\bftheta$ do not violate the imposition of the average kinematic variables from the macroscale (see \cite{glaesener_continuum_2019}). \eqref{eq:Wstar} defines a generally nonlinear optimization problem to be solved numerically in an iterative fashion\footnote{Although this minimization must be carried out numerically in general, all first and second derivatives with respect to the kinematic variables can be computed in closed-form, thus allowing for the efficient use of consistent tangents~\citep{glaesener_continuum_2019}.}. With the above effective energy density $W^*$ (dependent on the local unit cell architecture $\boldsymbol{\psi}$), we may employ principles of continuum mechanics to derive, e.g., the first Piola-Kirchhoff stress tensor as $\bfP=\partial W^*/\partial \bfF$ as well as analogous couple stress tensors associated with the remaining macroscale fields, as well as consistent stiffness matrices such as $\dsC=\partial^2 W^*/\partial\bfF\,\partial\bfF$.

Finally, for a given field $\bfpsi(\bfx)$ defining the local truss architecture, we can solve the macroscale boundary value problem on $\Omega$, viz.
\begin{equation}
\label{eq:VariationalProblem}
\begin{split}
	(\bfvarphi,\bftheta)_{\bfpsi} & = \arg \inf \left\{
	\int_\Omega W^*(\boldsymbol{\psi};\bfF,\bftheta,\bfkappa)\,\dd V
		- \int_{\partial\Omega_T} \hat\bfT \cdot\bfvarphi \dd S \right. \\
	& \qquad\qquad\qquad\qquad \left.	- \int_{\partial\Omega_m} \hat\bfm \cdot\bftheta \dd S
		\ : \ 
		\bfvarphi=\hat\bfvarphi\ \text{on}\ \partial\Omega_u, \  \bftheta=\hat\bftheta\ \textrm{on}\ \partial\Omega_\theta
		\right\}
\end{split}
\end{equation}
with tractions $\hat\bfT$ acting on the boundary $\partial\Omega_T\subset\partial\Omega$, and moments $\hat\bfm$ imposed on $\partial\Omega_m\subset\partial\Omega$. For conciseness, we introduce the generalized deformation mapping $\tilde{\bfvarphi}(\bfX)=(\bfvarphi(\bfX), \bftheta(\bfX))$ along with $\tilde{\bfF}=\bfnabla \tilde{\varphi}$ and $\hat\bft=(\hat\bfT,\hat\bfm)$, such that \eqref{eq:VariationalProblem} simplifies to
\begin{equation}
\label{eq:VariationalProblemSimplified}
\tilde{\bfvarphi} = \arg \inf \left\{ \int_\Omega W^*(\boldsymbol{\psi};\tilde{\bfvarphi},\tilde{\bfF})\,\dd V  - \int_{\partial\Omega_N} \hat\bft \cdot\tilde{\bfvarphi} \dd S:\tilde{\bfvarphi}=\hat{\bfvarphi} \textrm{ on } \partial\Omega_D \right\}
\end{equation}
with the combined Dirichlet boundary $\partial\Omega_D$.

\subsection{Topology optimization}
\label{subsec:topologyoptimzation}

In topology optimization we classically seek to find the optimal material distribution that minimizes an objective function under given constraints. Here, the global design field is $\boldsymbol{\psi}(\bfX)$, which we define to be varying continuously across $\Omega$ (or its finite element discretization $\Omega_h$). In our examples, we focus on classical benchmarks that minimize the compliance for single-load-case problems while enforcing a total volume (or mass) constraint. However, the framework presented here is sufficiently general to be extended to more complex objective functions and, specifically, it is already laid out for nonlinear problems (since the homogenization scheme accounts for finite beam rotations).

The optimization problem is formalized by 
\begin{equation}
\label{eq:toproblem}
\begin{split}
\min_{\boldsymbol{\psi}}: & \quad \calL(\boldsymbol{\psi}; \tilde{\bfvarphi}) 
= \int_\Omega \ell(\boldsymbol{\psi}; \tilde{\bfvarphi}) \dd V
= \int_\Omega \left[W^*(\boldsymbol{\psi};\tilde{\bfvarphi},\tilde{\bfF})+c_p || \bfnabla\psi_i ||_p\right] \dd V \\
\textrm{s.t.:} & \quad \int_\Omega W^*(\boldsymbol{\psi};\tilde{\bfvarphi},\tilde{\bfF})\,\dd V
= \int_{\partial\Omega_S} \hat\bft \cdot\tilde{\bfvarphi} \dd S, \\
: & \quad \int_\Omega \rho(\boldsymbol{\psi}) \dd V - V_{\textrm{max}} \leq 0,\\
: & \quad \text{RUC topology constraints},
\end{split}
\end{equation}
where $||(*)||_p$ denotes the $\bfL^p$ norm (in the following, we use $p=2$), and the $\bfL^p$-constraint \citep{borrvall_topology_2001}  governs the smoothness of the global design variable field $\bfpsi(\bfX)$ through a regularization constant $c_p \geq 0$.
$V_{\textrm{max}}$ defines the maximum material volume in the domain $\Omega$ (in subsequent examples we choose $\rho^\Omega=V_{\textrm{max}}/|\Omega| =0.2$), and the ``RUC topology constraints'' stand for any constraints imposed on the local truss topology and architecture (such as constraints on the unit cell topology, beam geometry, etc.). For the design updates of $\boldsymbol{\psi}$ we use the method of moving asymptotes \citep{svanberg_method_1987}.
Due to its nonlinear nature, the above optimization problem is solved iteratively, minimizing the objective with respect to the design variables $\boldsymbol{\psi}$ in a step-wise fashion. At each iteration step, the multiscale finite element problem defined in \eqref{subsec:macromodel} is solved for $\tilde\varphi$ in equilibrium. Computing the increments of the design field $\boldsymbol{\psi}$ requires computing its gradient, which is classically termed the sensitivity and given by 
\begin{equation}
    \frac{\partial \ell}{\partial\psi_i} = \frac{\partial W^*}{\partial \psi_i} + c_p \frac{\bfnabla \psi_i | \bfnabla \psi_i|^{p-2}}{||\bfnabla\psi_j||^{p-1}_p},
\end{equation}
where the semi-analytical homogenization scheme introduced in Section~\ref{subsec:macromodel} allows us to efficiently calculate the change of the condensed energy density $\partial W^*$, locally in each quadrature point, with respect to a design variable change $\partial \psi_i$ through a local forward-difference scheme (introducing a small perturbation on each design variable of $10^{-9}$). 

\begin{figure}
\centering
\begin{subfigure}[t]{.45\textwidth}
  \centering
        \includegraphics[width=\linewidth]{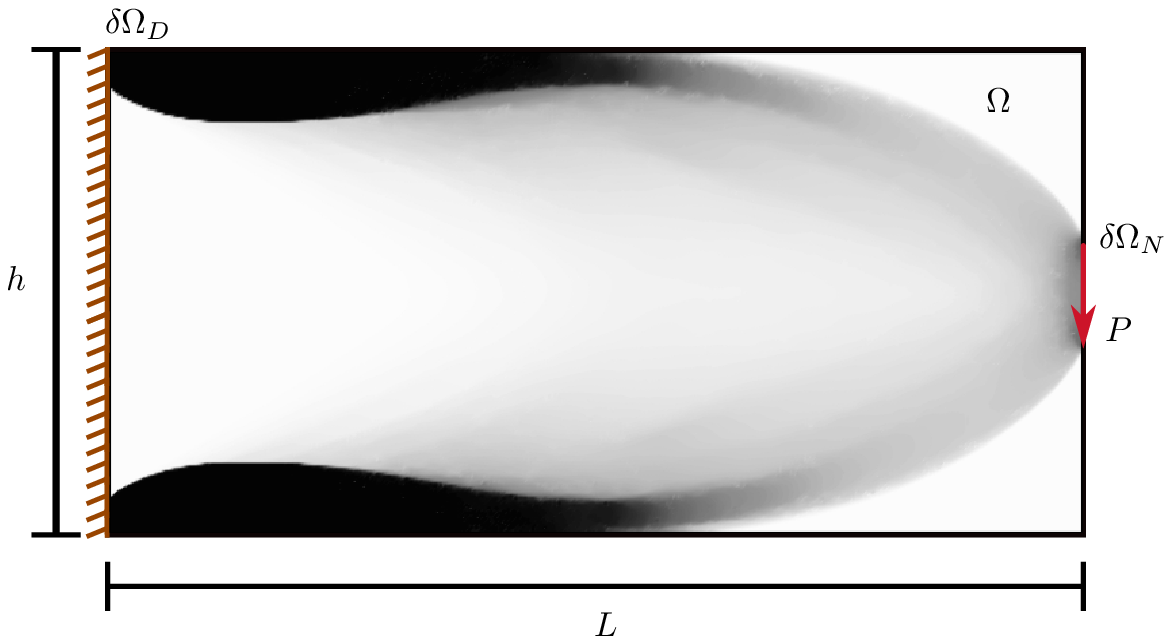}
  \caption{Continuum cantilever beam problem as defined in \citep{sigmund_non-optimality_2016}. All nodes on the left edge are clamped. The load $P=10^{-6}$ is applied as distributed load over the central $20\%$ of the right edge. Solid black indicates $50\%$ volume fraction.}
  \label{fig:continuumCantileverBeam}
\end{subfigure} \qquad
\begin{subfigure}[t]{.45\textwidth}
  \centering
        \includegraphics[width=\linewidth]{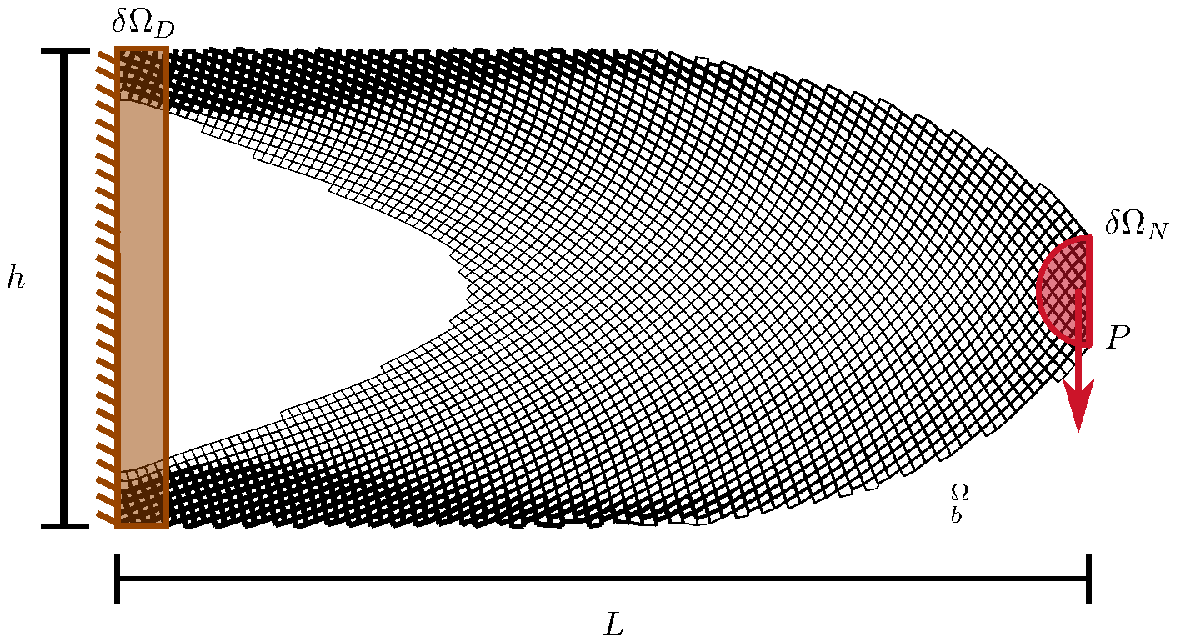}
  \caption{Discrete cantilever beam. All nodes with distance to the left edge smaller than $0.1h$  are clamped. The load $P$ is distributed to all nodes with distance smaller $0.1h$ to the centerright point. This image serves for visualization purposes only.}
  \label{fig:discreteCantileverBeam}
\end{subfigure}
\caption{The cantilever beam is well studied and serves as benchmark. In this study, the beam domain $\Omega$ has a length $L=2$ and a height $h=1$. Furthermore, we choose the load $P=10^{-6}$.  }
\label{fig:cantileverBeam}
\end{figure}

For a performance comparison to literature benchmarks, we restrict our study to the cantilever beam in \autoref{fig:continuumCantileverBeam} with an applied load of magnitude $P$ and define the beam compliance, normalized by the effective density $\rho^\Omega$, as 
\begin{equation}
\label{eq:ComplianceCont}
\calC_C(\tilde\bfx) = \frac{\rho^\Omega}{P^2}  \int_\Omega W^*(\boldsymbol{\psi};\tilde{\bfvarphi},\tilde{\bfF}) \dd V.
\end{equation}

In subsequent examples, we study three different RUC families in $d=2$ dimensions (2D), whose truss architectures are constrained by the RUC topology constraints in \autoref{tab:RUCconstraints}. The square lattice family resembles the study in \cite{groen_homogenization-based_2018} and serves as a benchmark. The rectangular lattice family, and ultimately the monoclinic Bravais lattice, exemplify two generalizations facilitated by this approach. Note that the effective density $\bar{\rho}$ is analogous to classical density-based topology optimization, while the Bravais lattice variables ($\alpha$, $\beta$, $\gamma$) allow for tailoring of the local anisotropy of the effective elastic surfaces. By letting the effective density vary up to $50\%$, we are stretching the assumption of slender beams. Please find a detailed discussion and reasoning of the upper density bound in \ref{sec:beam_slenderness}. As initial guess we use a homogeneous design variable field with values in the center of the design variable ranges.

\begin{table}[h]
\caption{RUC topology constraints for tetragonal and orthorhombic crystal lattice families, and a ``monoclinic'' Bravais lattice parametrization that includes all primitive Bravais lattice families in 2D (tetragonal, hexagonal, orthorhombic, monoclinic). While the tetragonal lattice family is widely used in the literature, other lattice families are oftentimes neglected, which restricts the design space.}
\label{tab:RUCconstraints}
\centering
\begin{tabular}{@{}ccc@{}}
\toprule
tetragonal & orthorhombic &  monoclinic \\ \midrule
$-\pi \leq \alpha \leq \pi$ & $-\pi \leq \alpha \leq \pi$ & $-\pi \leq \alpha \leq \pi$                         \\ 
$\beta = \pi/2$ & $\beta = \pi/2$      & $0 \leq \beta \leq \pi$        \\ 
$\gamma = 1$            & $0.5 \leq \gamma \leq 2$ & $0.5 \leq \gamma \leq 2$ \\
$10^{-5} \leq \bar{\rho} \leq 0.5 $ & $10^{-5} \leq \bar{\rho} \leq 0.5 $ & $10^{-5} \leq \bar{\rho} \leq 0.5 $             \\ \bottomrule
\end{tabular}
\end{table}

\subsection{From the continuous solution to a discrete structure}
\label{subsec:continuoustodiscrete}

Solving the optimization problem \eqref{eq:toproblem} results in a continuous distribution of the truss design parameter field $\bfpsi$ across $\Omega$ on the macroscale, whose underlying microstructure is, in principle, infinitely fine based on the separation of scales.
In order to construct a discrete truss at a finite length scale (based on the continuous optimal field $\bfpsi$ obtained from \eqref{eq:toproblem}), we here extend the approach presented in \cite{rumpf_synthesis_2012}.

We introduce a mapping $\epsilon(\bfX):\Omega\to[0,1]$, such that a value of $0$ refers to one phase (e.g., void) and a value of $1$ to another phase (e.g., solid). Most generally, we represent $\epsilon(\bfX)$ of a periodic unit cell in terms of a Fourier series (e.g., we find the complex amplitudes for a discretely defined unit cell through a fast Fourier transform). When assuming translational symmetry such as in a lattice, we use the symmetry information and represent the infinite structure in terms of the Fourier series over reciprocal lattice vectors, most generally written as
\begin{equation}\label{eq:MappingFourier}
    \epsilon(\bfX) = \sum_{\bfk} b_{\bfk} e^{i 2 \pi \mu \bfk \cdot \bfX} \quad \textrm{with} \quad \bfk=\sum_{i=1}^d m_i\bfk_i
\end{equation}
being an integer combination of the primitive reciprocal lattice vectors $\bfk_i$, and $b_{\bfk}$ denoting the complex amplitude related to the spatial harmonic with reciprocal lattice vector $\bfk$. $\mu$ represents a dimensionless length scale introduced to govern the unit cell size. Translational symmetry is ensured by the primitive reciprocal lattice vectors which, in $\calR^3$, are given by
\begin{equation}\label{eq:reciprocallatticevectors3D}
    \bfk_1= \frac{\bfa_2 \times \bfa_3}{\bfa_1 \cdot (\bfa_2 \times \bfa_3)}, \qquad
    \bfk_2= \frac{\bfa_3 \times \bfa_1}{\bfa_2 \cdot (\bfa_3 \times \bfa_1)}, \qquad
    \bfk_3= \frac{\bfa_1 \times \bfa_2}{\bfa_3 \cdot (\bfa_1 \times \bfa_2)},
\end{equation}
based on the primitive Bravais lattice vectors $\bfa_i$. In 2D, the reciprocal lattice vectors $\bfk_1$ and $\bfk_2$ follow from \eqref{eq:reciprocallatticevectors3D} with $\bfa_3=(0,0,1)$.

When constructing a spatially variant truss, the spatial variation of $\bfk(\bfX)$ renders \eqref{eq:MappingFourier} problematic, since the latter requires a constant history of the frequency vectors $\bfk$ in space.
When violated, highly distorted projections arise. As a remedy, in \cite{rumpf_synthesis_2012}  a projection function $\phi_{\bfk}(\bfX)$ was introduced to obtain the mapping 
\begin{equation}
    \epsilon^\prime(\bfX) = \sum_{\bfk} b_{\bfk} e^{i 2 \pi \mu \phi_{\bfk}(\bfX)} 
\end{equation}
with lattice symmetry enforced locally through
\begin{equation}
\label{eq:ProjectionFunction}
    \nabla \phi_{\bfk}(\bfX) = \bfk(\bfX).
\end{equation}
That is, we ensure that $\nabla\phi_{\bfk}$ is locally tangential to the (local) $\bfk$-vector defined through \eqref{eq:reciprocallatticevectors3D} at every point on the macroscale.

Finding the field $\phi:\Omega\to\Rset^d$, in principle, requires the solution of an over-determined system. When the lattice is perfectly periodic, the system has the solution $\phi_{\bfk}(\bfX)=\bfk\cdot\bfX+c$ with arbitrary $c\in\Rset$. By contrast, for all spatially variant trusses, we find $\phi_{\bfk}$ as a least-square minimizer:
\begin{equation}
    \label{eq:leastSquares}
    \phi_{\bfk}(\bfX) = \arg\min\int_\Omega ||\nabla\phi_{\bfk}(\bfX)- \bfk ||^2\dd\Omega.
\end{equation}
Given a separation of scales between the spatial variation of the lattice vectors and the introduced RVE length scale $\mu$, this ensures quasi-periodicity in the neighborhood of each RVE (on the microscale). The global residual defined by \eqref{eq:leastSquares} (which is generally non-zero) can thus be interpreted as a measure of the violation of scale separation.

\begin{figure}
	\centering
	\begin{subfigure}[t]{0.33\linewidth}
	    \centering
 		\includegraphics[page=1,width=.75\linewidth]{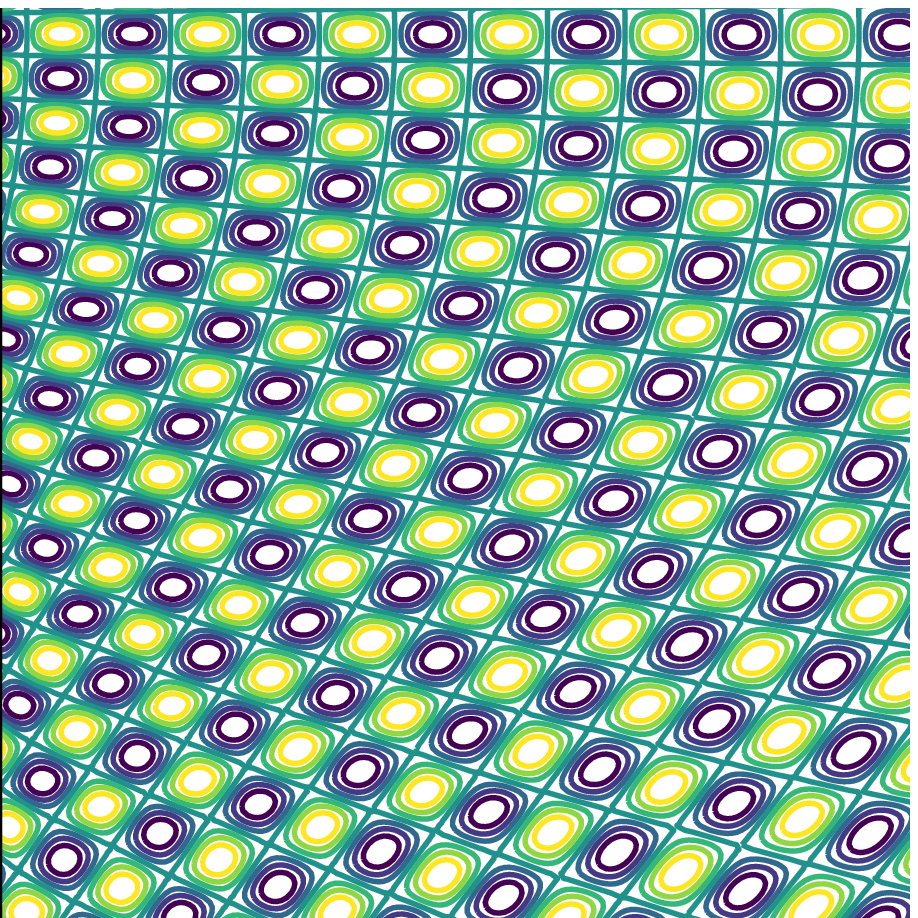}
		\caption{lattice mapping function $\tau^\prime(\bfX)$}
		\label{fig:lattice_function}
	\end{subfigure}%
	\hfill
	\begin{subfigure}[t]{0.33\linewidth}
	    \centering
 		\includegraphics[page=1,width=.75\linewidth]{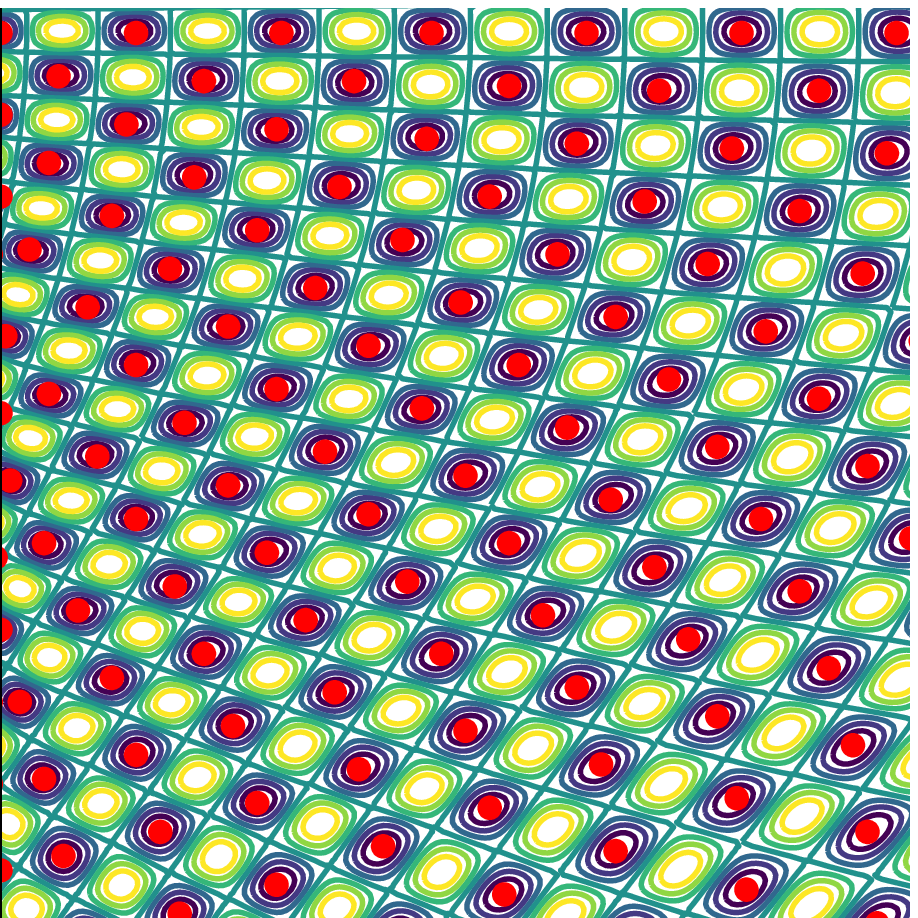}
		\caption{lattice sites}
		\label{fig:projection_minima}
	\end{subfigure}%
	\hfill
	\begin{subfigure}[t]{0.33\linewidth}
	    \centering
 		\includegraphics[page=1,width=.75\linewidth]{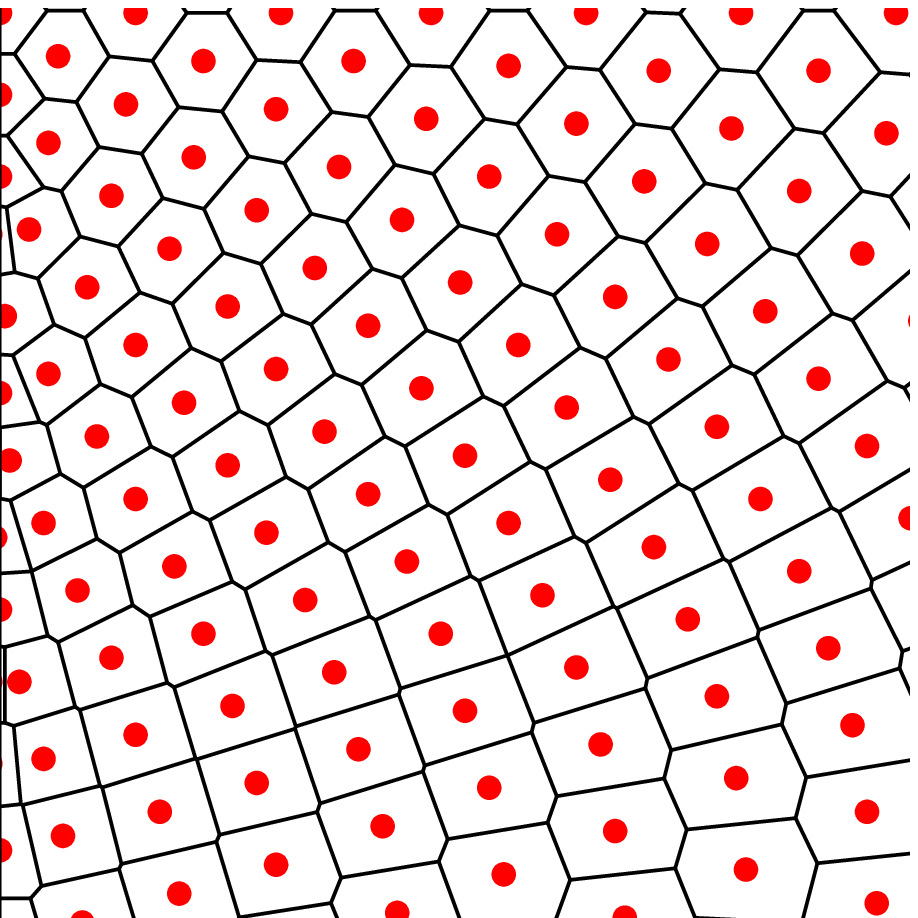}
		\caption{Voronoi diagram}
		\label{fig:voronoi}
	\end{subfigure}%
	\caption{The discrete lattice is constructed from the lattice mapping function by first identifying the lattice sites, e.g., at lattice mapping function minima, and second constructing the Voronoi diagram over all lattice sites.}
	\label{fig:projection}
\end{figure}

For the lattice description introduced in Section~\ref{subsec:spatiallyvariantlattices}, we define a mapping function $\tau(\bfX)$ which attains a value of $1$ at each lattice site while being $0$ elsewhere:
\begin{equation}
\label{eq:periodic_voronoi_mapping}
\tau(\bfX) 
=
\prod_{i=1}^d \sum_{n=-\infty}^\infty \delta \left( 
\mu\frac{\bfk_i\cdot\bfX}{||\bfk_i||} - n||\bfk_i|| 
 \right)
=
\prod_{i=1}^d  \Sha \left(\mu\bfk_i\cdot\bfX \right),
\end{equation}
where $\delta(x)$ and $\Sha(x)$ denote the Kronecker delta and Kronecker delta impulse train, respectively, in $d$ dimensions. Decomposing \eqref{eq:periodic_voronoi_mapping} into spatial harmonics, the lattice mapping function can be represented as the Fourier series
\begin{equation}
\label{eq:periodic_voronoi_mapping_fourier}
    \tau(\bfX) =
    \prod_{i=1}^d \frac{1}{1+2M} \sum_{m=-M}^M e^{i 2 \pi m \mu \bfk_i \cdot \bfX }  = 
    \prod_{i=1}^d \frac{1}{1+M} \left[ 1 + \sum_{m=1}^M \cos(2  \pi m \mu \bfk_i \cdot \bfX ) \right].
\end{equation}
Note that, in general, an infinite number ($M\to\infty$) of spatial harmonics is required to represent the Dirac combs exactly. For implementation purposes, we truncate the above Fourier series and, using the fact that the translational symmetry is contained in all spatial harmonics, choose $M=1$ such that
\begin{equation}
\label{eq:truncated_voronoi_mapping}
    \tau^\prime(\bfX) = \frac{1}{2^d} \prod_{i=1}^d  \left[1 + \cos(2  \pi \mu \phi_{\bfk_i}\big(\bfX)\big)\right]
\end{equation}
represents the truncated lattice mapping function with locally enforced lattice symmetry according to \eqref{eq:ProjectionFunction}. We thus converted the strictly periodic and discontinuous lattice mapping \eqref{eq:periodic_voronoi_mapping} into a spatially varying and $\calC^\infty$-continuous truncated mapping function \eqref{eq:truncated_voronoi_mapping}, whose local maxima represent lattice site locations and have value $1$ for convenience of implementation.
We thus identify the lattice sites within $\Omega$ as the local maxima of \eqref{eq:truncated_voronoi_mapping} (see \autoref{fig:projection_minima}). Finally, the truss is constructed by extracting all nodes $\calB^\Omega_n$ and beams $\calB^\Omega_b$ of width $w(\bfX)>w_\text{min}$ -- analogous to the construction of the RUC described in Section~\ref{subsec:spatiallyvariantlattices} -- from the Voronoi diagram over all lattice sites in the global domain $\Omega$ (see \autoref{fig:voronoi}). The threshold $w_\text{min}$ reflects the lower bound on $\bar{\rho}$ set in \autoref{tab:RUCconstraints}.

\subsection{Discrete description}
\label{subsec:discretediscription}

With the effective continuum description introduced in Section \ref{subsec:macromodel} we efficiently approximate the lattice behavior, as demonstrated previously for periodic lattices in \cite{glaesener_continuum_2019}. To validate its applicability to the spatially-variant lattices introduced here, we compare to full discrete simulations.

Based on the global sets of nodes and beams, $\calB^\Omega_n$ and $\calB^\Omega_b$, respectively (see Section \ref{subsec:spatiallyvariantlattices}), we define the total potential energy of the beam network subjected to generalized external nodal tractions $\hat \bft$ acting on all nodes in $\calB^{\partial \Omega_N}_n$ as
\begin{equation}
\label{eq:PotentialEnergyDisc}
    I(\tilde\bfx) = 
     \sum_{\nu \in \calB^\Omega_b} W_\nu(\tilde{\bfx}_{\nu_1}, \tilde{\bfx}_{\nu_2}) -
     \sum_{\alpha \in \calB^{\partial\Omega_N}_n} (\hat \bft_\alpha \cdot \tilde \bfx_\alpha ),
\end{equation}
where $\tilde \bfx = \{ \tilde \bfx_1, \dots, \tilde\bfx_{n_n}\}$ is the set of all nodal degrees of freedom. Unlike for the continuum cantilever beam in \autoref{fig:continuumCantileverBeam}, the load $P$ is distributed over a semi-circular area $\partial\Omega_N$, as shown in \autoref{fig:discreteCantileverBeam}, to assure an even load distribution between discrete nodes. For analogous reasons, the fixed boundary condition at the left boundary of the domain is spread over a thin slice $\calB^{\partial \Omega_D}_n$. Classically, the generalized force vector of node $\alpha$, containing both forces and moments, and the associated stiffness matrix follow as, respectively,
\begin{equation}
\label{eq:TractionsDisc}
    \tilde \bft_\alpha (\tilde \bfx) = \frac{\partial I}{\partial \tilde \bfx_\alpha}(\tilde \bfx), \qquad \tilde \bfK_{\alpha \beta} (\tilde \bfx) = \frac{\partial I}{\partial \tilde \bfx_\alpha \partial \tilde \bfx_\beta}(\tilde \bfx).
\end{equation}
We obtain the quasistatic response from energy minimization, i.e.,
\begin{equation}
\label{eq:MinimizationDisc}
    \tilde \bfx = \arg \min \{I:\tilde \bfx_\alpha = \bf0 \ \text{for} \ \alpha \in \calB^{\partial \Omega_D}_n\},
\end{equation}
where the generalized displacements $\tilde \bfx_\alpha$ of all nodes in $\calB^{\partial \Omega_D}_n$ near the left boundary are enforced to be zero. For comparability with literature and the compliance of the effective continuum description \eqref{eq:ComplianceCont}, we define the normalized compliance for the discrete beam network as
\begin{equation}
\label{eq:ComplianceDisc}
\calC_D(\tilde\bfx) = \frac{\rho^\Omega}{P^2} \sum_{\nu \in \calB^\Omega_b} W_\nu(\tilde{\bfx}_{\nu_1}, \tilde{\bfx}_{\nu_2}), \qquad \text{where} \qquad \rho^\Omega= \frac{1}{|\Omega|}\sum_{\nu \in \calB^\Omega_b} w_\nu L_\nu t \rho
\end{equation}
with $L_\nu$ denoting the length of beam $\nu$.

\section{Numerical implementation}
\label{sec:numerical_implementation}
The mechanical boundary value problem on the macroscale, described by the variational optimization problem \eqref{eq:VariationalProblemSimplified} or \eqref{eq:MinimizationDisc}, is solved via a finite-element approximation which discretizes $\Omega$ into $\Omega_h$, using bi-linear quadrilateral elements to result in bi-linear interpolations of both the displacement field $\bfvarphi_h$ and the rotation field $\bftheta_h$ across $\Omega_h$. To accommodate the micropolar nature of the discretization, we use an in-house finite-element code, which solves the stationarity conditions (i.e., the equations of linear momentum balance) via Newton-Raphson iteration for quasistatic equilibrium. Further details about the numerical treatment of the homogenized continuum model can be found in \cite{glaesener_continuum_2019}.

The design parameter field $\psi(\bfX)$, defined in \eqref{eq:designparams}, is interpolated using the exact same finite element approximation based on bi-linear quadrilateral elements, thus providing a continuous field of spatially varying truss unit cell architectures. As pointed out in Section~\ref{subsec:topologyoptimzation}, finding the solution for $\psi(\bfX)$ involves the method of moving asymptotes. The overall optimization problem is solved in an iterative fashion, minimizing the objective function with respect to field $\psi(\bfx)$, while solving the macroscale boundary value problem.

To construct the discrete truss in a postprocessing step from the given interpolated $\psi$-field, we first solve for the projection function \eqref{eq:leastSquares}, using again the same finite element approximation, and then identify the lattice sites as local maxima of \eqref{eq:truncated_voronoi_mapping}. Only at this step we introduce the finite length scale $\mu$. Finally, we build the global Voronoi diagram and extract node locations and beam connections.

\section{Numerical examples}
\label{sec:numerical_examples}
The introduced methodology is a tool for the optimization of generic structures based on a free choice of the underlying Bravais RUC topology and constituents. In the following, we showcase the potential of this approach by studying the optimization of tetragonal, orthorhombic and monoclinic truss lattices, and we investigate the influence of the RUC constraints. Moreover, we discuss the underlying assumption of a separation of scales and demonstrate how it can be (approximately) fulfilled in the lattice projection as a trade-off between continuum design variable field regularization and choice of length scale.

\subsection{Tetragonal lattice family}

The tetragonal truss lattice family (see \autoref{tab:RUCconstraints}) has been extensively studied in the literature and is the basis of various lattice optimization schemes (e.g. \cite{khanoki_multiscale_2012, zhu_two-scale_2017, groen_homogenization-based_2018}). Hence, the tetragonal lattice family serves as an excellent introductory example and benchmark for our method. We first present optimized continuum design variable results and discuss the influence of mesh element size and the relaxation constant. Afterwards, we project the continuous fields onto a finite scale and compare the performance of the discrete truss lattice  with that of the optimized continuum.

\subsubsection*{Continuum}
\label{Sec:TetragonalContinuum}

\begin{figure}[!b]
	\centering
	\begin{subfigure}[t]{0.26\linewidth}
    	\centering
     	\includegraphics[angle=90,width=\linewidth]{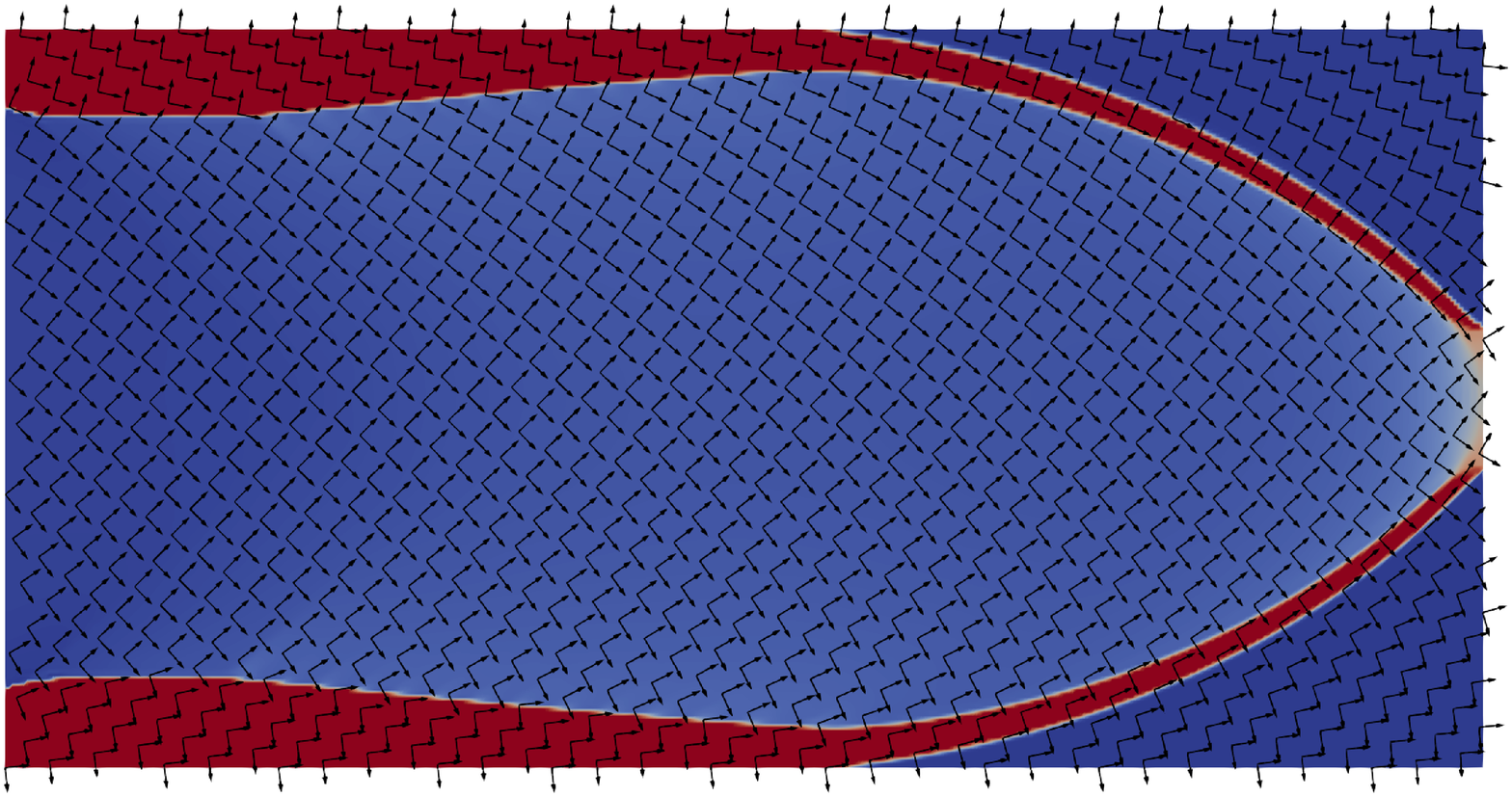}
    	\caption{$c_p=10^{-13}$}
    	\label{fig:Square_400x200_13}
	\end{subfigure}
	\hfill
	\begin{subfigure}[t]{0.26\linewidth}
    	\centering
     	\includegraphics[angle=90,width=\linewidth]{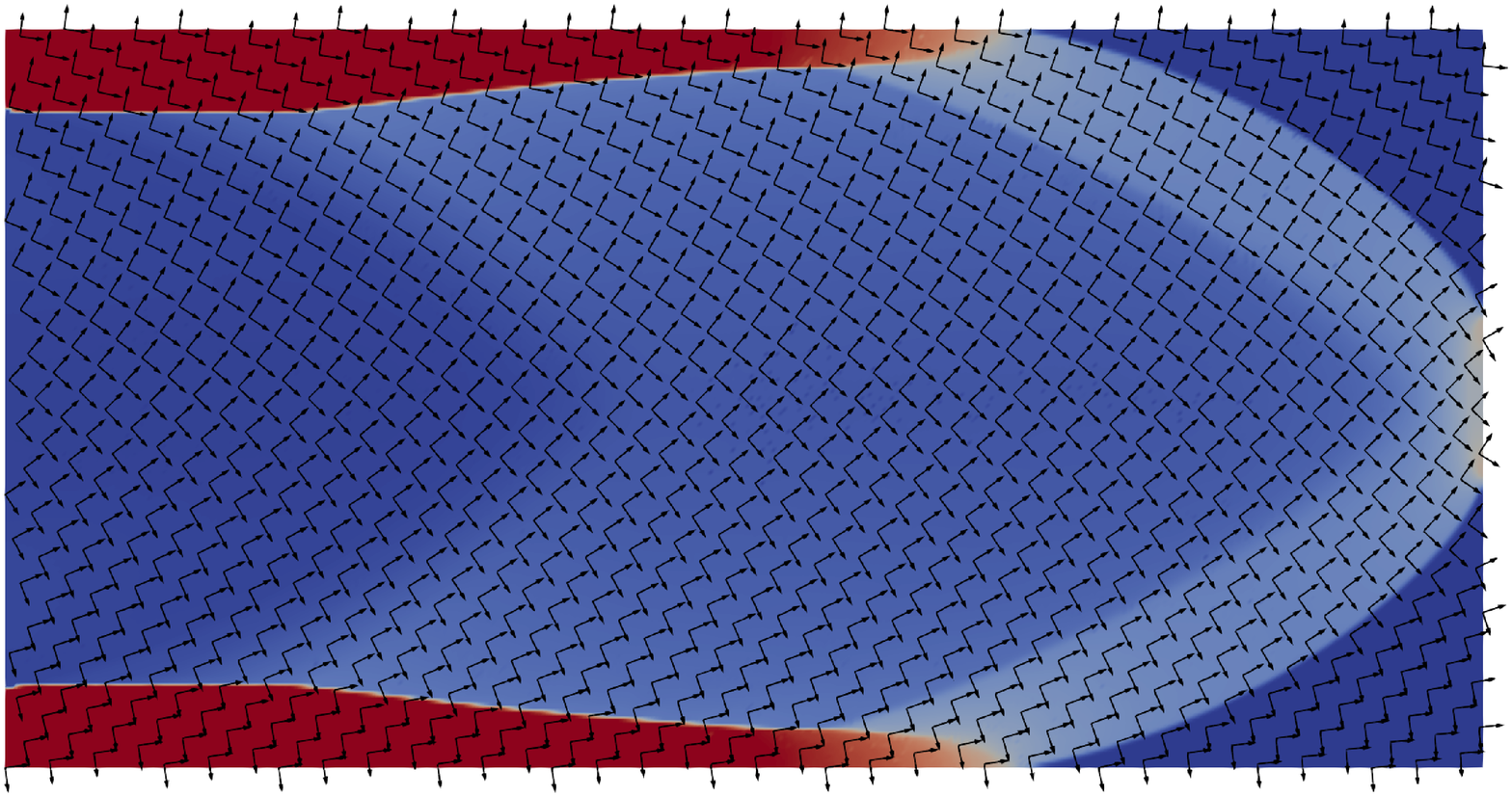}
    	\caption{$c_p=10^{-12}$}
    	\label{fig:Square_400x200_12}
	\end{subfigure}
	\hfill
	\begin{subfigure}[t]{0.26\linewidth}
    	\centering
     	\includegraphics[angle=90,width=\linewidth]{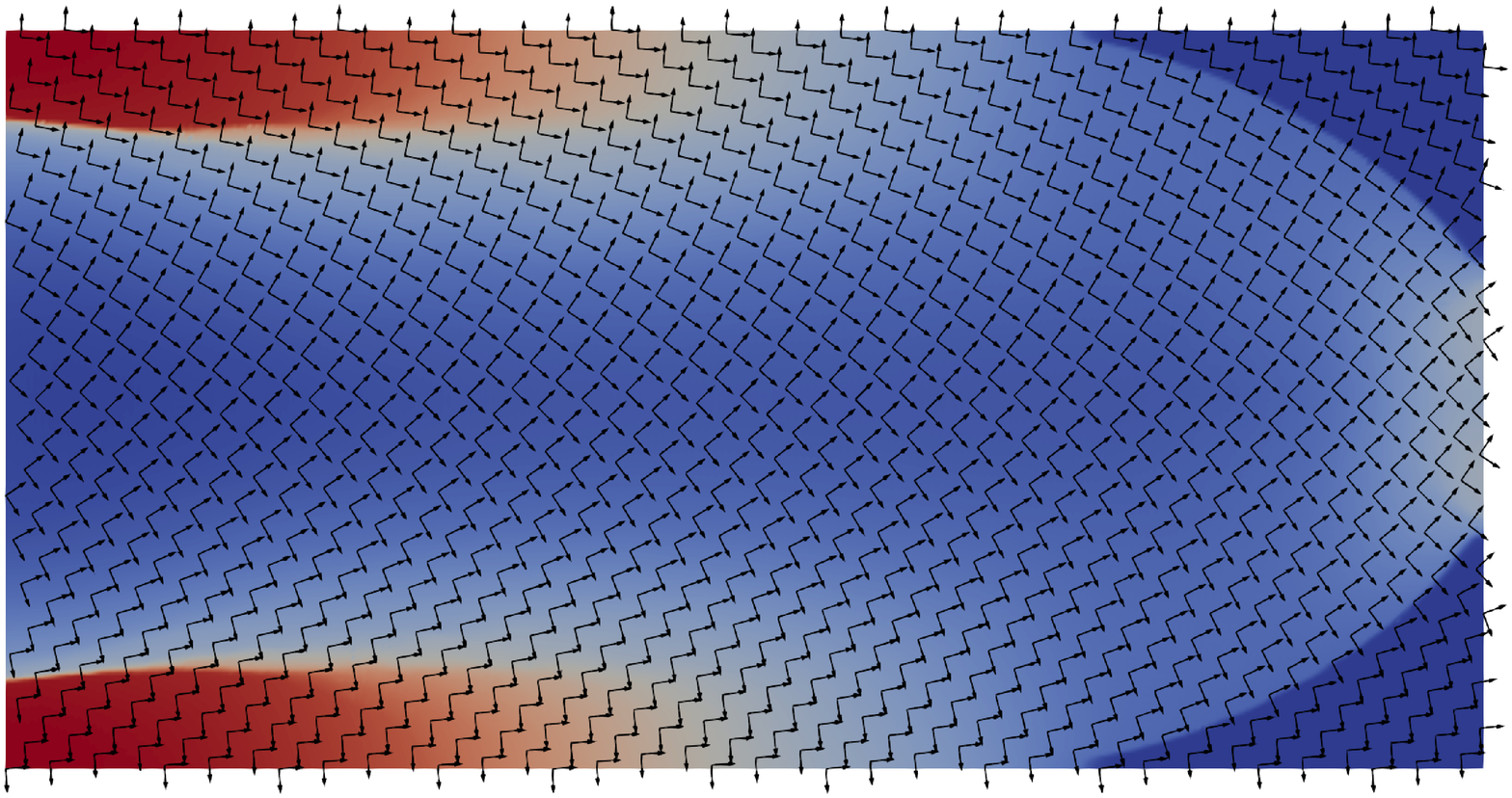}
    	\caption{$c_p=10^{-11}$}
    	\label{fig:Square_400x200_11}
	\end{subfigure}
	\hfill
	\begin{subfigure}[t]{0.05\linewidth}
    	\centering
     	\includegraphics{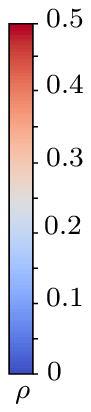}
	\end{subfigure}
	\caption{Optimized cantilever beam based on a tetragonal truss lattice for different regularizations. Local Bravais bases are indicated by arrows, and the density is shown by the color code. The arrows are scaled and shown at selected locations for readability. Resolution: 400x200 elements. All plots are rotated by 90$^\circ$. Full design variable fields can be found in \autoref{fig:SquareContinuumFields}.}
	\label{fig:Square_400x200}
\end{figure}

\begin{table}[!b]
\caption{Minimized continuum compliance $\calC_\text{C}$ for tetragonal, orthorhombic, and monoclinic crystal lattice families with different regularization constants $c_p$ and initial guesses. The (H)omogeneous initial guess assumes initial angles $\alpha=\pi/4$, $\beta=\pi/2$, $\gamma=1$, and $\rho=\rho^\Omega=0.2$. The (O)rthotropic initial guess uses the optimized fields from the orthotropic lattice family as initial guess.}
\label{tab:MinimizedContinuumCompliance}
\centering
\begin{tabular}{ccccccc}
\toprule
	&	tetragonal	&\phantom{a}&	orthorhombic	&\phantom{a}& \multicolumn{2}{c}{monoclinic} \\ \cmidrule{2-2}\cmidrule{4-4}\cmidrule{6-7}
$c_p$	&	(H)	&&	(H)	&& (H)	& (O) \\ \midrule
$10^{-13}$  & 	66.6	&&	62.2	&&	76.3	&	62.2  \\
$10^{-12}$  & 	67.4	&&	63.3	&&	78.4	&	63.3  \\ 
$10^{-11}$	& 	84.8	&&	79.9	&&	88.2	& 	79.9  \\ \bottomrule
\end{tabular}
\end{table}
              
\begin{figure}
	\centering
		\centering
     	\includegraphics{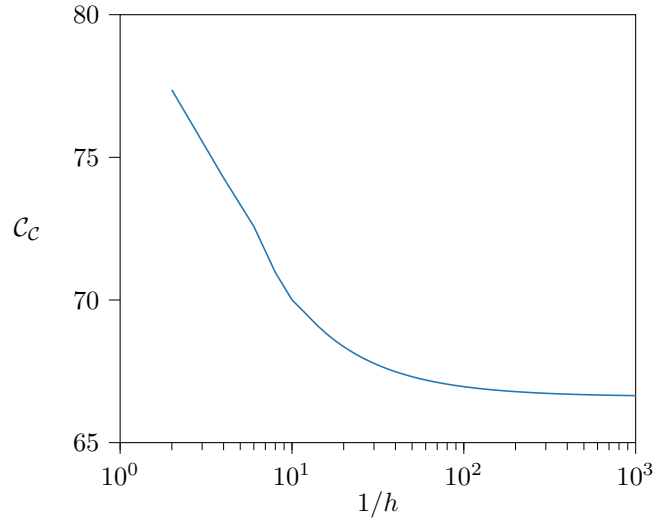}
		\caption{Compliance versus mesh resolution for tetragonal lattice cantilever problem. A mesh of 80x40 elements is within $2\%$ of the converged compliance $66.6$. $c_p=10^{-13}$.}
		\label{fig:hConvergence}

\end{figure}

We solve the optimization problem stated in \eqref{eq:toproblem}, subjected to the tetragonal RUC constraints (see \autoref{tab:RUCconstraints}), on a continuum mesh of 400x200 Q4 elements. The mesh resolution was chosen as a trade-off between convergence (see \autoref{fig:hConvergence}), fine resolution for projection step, and acceptable computation time ($<20$ minutes on a regular laptop until convergence). The infimizing design variable fields $\bfpsi(\bfX)$ for low ($c_p=10^{-13}$), medium ($c_p=10^{-12}$) and strong ($c_p=10^{-11}$) regularization are presented in \autoref{fig:SquareContinuumFields}. The resulting Bravais bases vectors and density field can be found in \autoref{fig:Square_400x200}.   The corresponding compliance values are shown in \autoref{tab:MinimizedContinuumCompliance}.

\begin{figure}
	\centering
	\begin{subfigure}[t]{0.48\linewidth}
		\centering
     	\includegraphics[width=\linewidth]{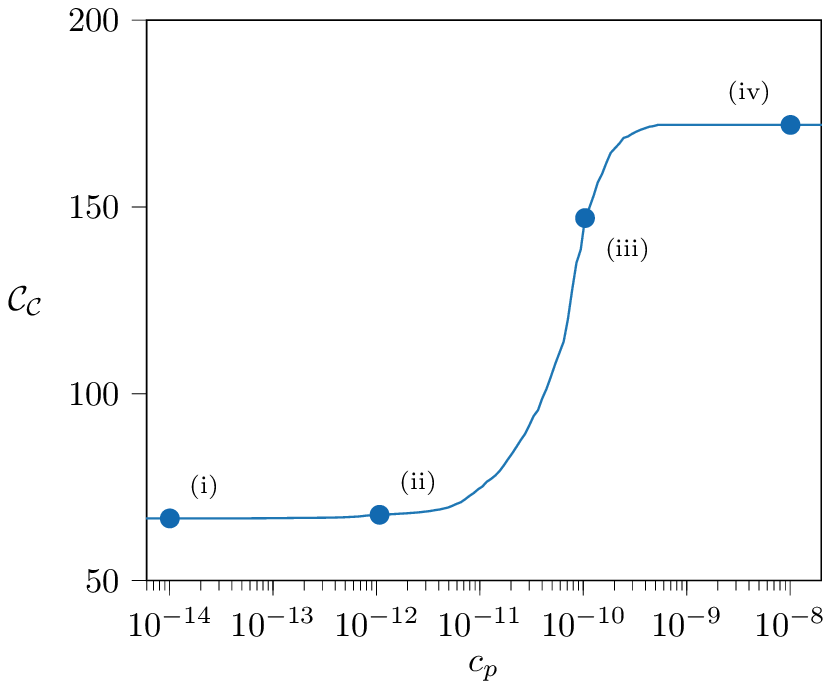}
		\caption{Compliance versus regularization constant. For a low regularization constant the compliance converges to a minimum, while for high regularization constants the compliance plateaus at its maximum.}
		\label{fig:filterInfluence_compliance}
	\end{subfigure}%
	\hfill
	\begin{subfigure}[t]{0.48\linewidth}
    	\centering
     	\includegraphics[width=\linewidth]{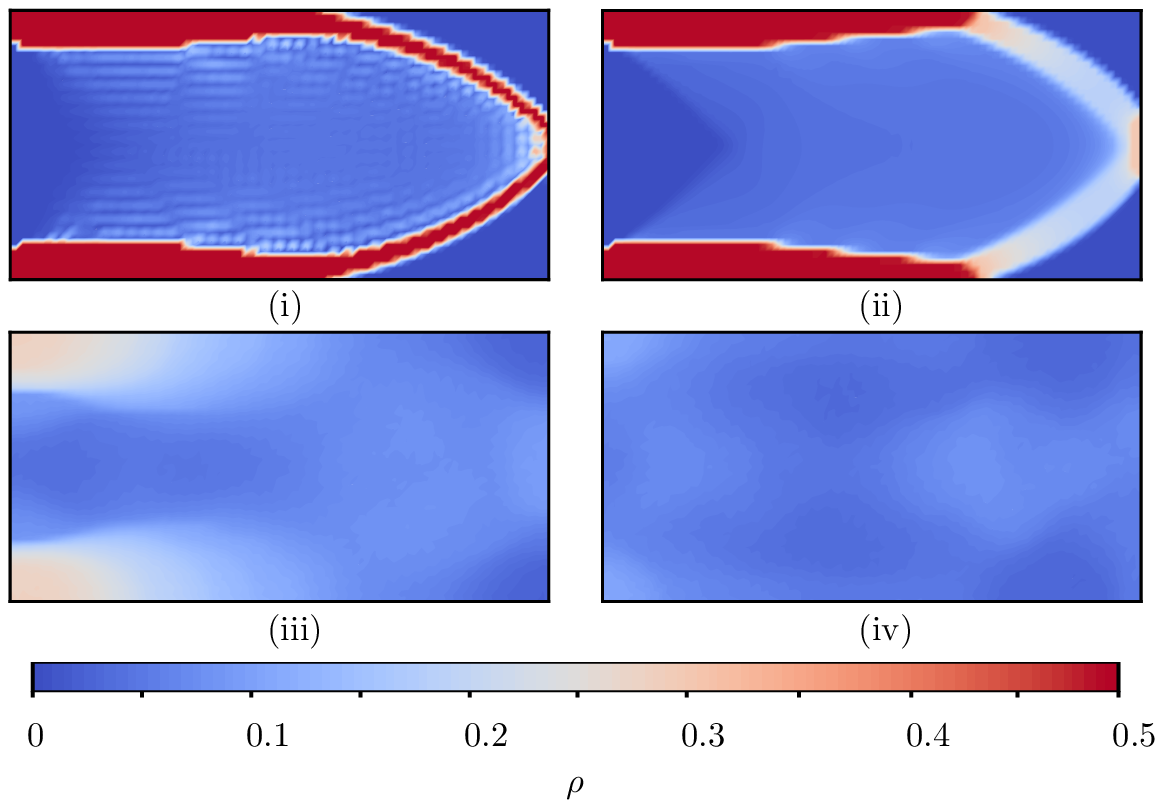}
    	\caption{Effective density for the optimized cantilever beam based on a tetragonal lattice for different relaxation constants $c_p$, as indicated by the markers in (a). Discretization: 80x40 elements.}
    	\label{fig:filterInfluence_density}
	\end{subfigure}
	\caption{Influence of the regularization constant $c_p$.}
	\label{fig:filterInfluence}
\end{figure}

The regularization, as a constraint on the design variable field, has a detrimental effect on the effective compliance. The regularization's influence on the compliance and density field is illustrated in \autoref{fig:filterInfluence_compliance} and \autoref{fig:filterInfluence_density}, respectively. While for low regularization, its influence on compliance and density is negligible (see case (i)), the compliance is affected with increasing $c_p$  (case (ii)). When further increasing, the compliance degrades strongly (case (iii)) and finally reaches a second plateau (case (iv)). This upper plateau refers to spatially constant values of $\alpha$ and $\rho$.  Given the negative effect on structural performance, one might wonder why we use regularization. The reasons are twofold.

First, with no or too little regularization, numerical (interpolation-dependent) artifact patterns arise (see case~(i)). The effect is comparable to the well-known checkerboarding in classical topology optimization with discrete density fields \citep{SigPet98}. The chosen FE discretization leads to artificially stiff results, and the projection of such laminated lattices onto a finite scale would result in disconnected lattices. Regularization alleviates this problem. Alternatively, the problem could be avoided by increasing the element interpolation order. However, we here choose the regularization to control the length scale and restrict our finite elements to linear interpolation for simplicity.

Second, the continuously described RUCs may feature abrupt changes of the design variables (e.g., when crossing finite element boundaries on the macroscale). When projecting those onto a finite discrete lattice, this may either locally break the  assumption of quasi-periodicity (see \autoref{eq:SpatiallyVariantBravaisLattice}) or require high resolution of the discrete projection (in fact, infinite resolution in the case of discrete design variable jumps, so $\mu \to 0$). The regularization constant gives us control over the locally permitted change of the unit cell in the lattice and hence guarantees a finite length scale ($\mu>0$) of the discrete lattice.

In comparison with optimization results of other topology optimization schemes, the compliance for the tetragonal lattice with regularizations $c_p=10^{-13}$ ($c_p=10^{-12}$, $c_p=10^{-11}$) is $34.9\%$ ($36.6\%$, $71.8\%$) higher than the analytical solution ($49.35$) in \cite{chan_design_1962} and $14.1\%$ ($15.5\%$, $45.3\%$) higher than the benchmark ($58.35$) in \cite{groen_homogenization-based_2018}. This is expected, as the topology space for the tetragonal unit cell description is the most limited of this study. It can reproduce the same unit cell topology as in  \cite{groen_homogenization-based_2018}, but it lacks the possibility to tune the directionality by adjusting the widths of each beam in a unit cell.
Moreover, in \cite{pedersen_optimal_1989} it was shown that the optimal orientation of an orthotropic material -- like the tetragonal lattice (see \autoref{fig:ElasticSurfaces}) -- aligns with the principal stress directions. Here, $\alpha$ governs the unit cell orientation and indeed aligns beams with the principle stress directions. One may argue that this knowledge could be used in the optimization process. However, when the other RUC parameters ($\beta$, $\gamma$, and $\bflambda$) are subject to optimization, $\alpha=0$ does not necessarily point in the direction of highest stiffness (see \autoref{fig:ElasticSurfaces}). In the following sections we will study the influence of changing orthotropy through changes of $\beta$ and $\gamma$.

\subsubsection*{Discrete lattice}
\label{Sec:TetragonalDiscrete}

\begin{figure}
\centering
\begin{minipage}[t]{0.48\textwidth}	
    \centering
    \includegraphics[width=\linewidth]{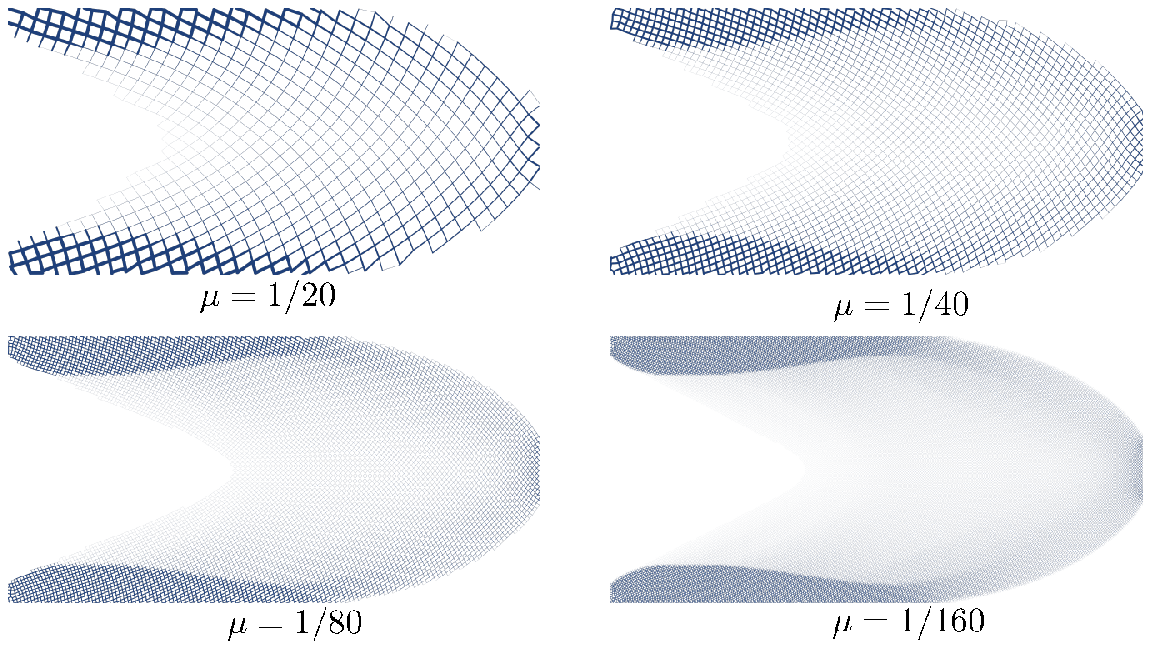}
    \caption{Tetragonal lattice cantilever beam at different length scales for $c_p=10^{-12}$. For large values of $\mu$, the structure resembles gradually the density field in \autoref{fig:SquareContinuumFields}. At $\mu=1/160$ the lattice consists of more than 120{,}000 elements.}
    \label{fig:TetragonalBeamMesh}
\end{minipage}%
\hfill
\begin{minipage}[t]{0.48\textwidth}	
	\centering
 	\includegraphics[width=\linewidth]{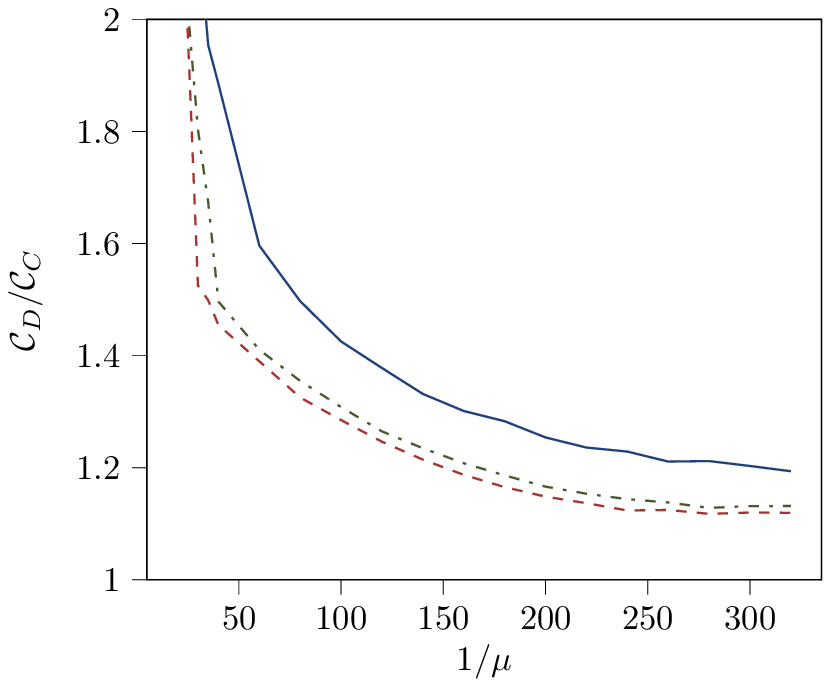}
	\caption{Convergence of the discrete compliance $\calC_D$ to the continuum approximation $\calC_C$ for regularization constants $c_p=10^{-13}$ (solid, blue), $c_p=10^{-12}$ (dash-dotted, green), and $c_p=10^{-11}$ (dashed, red).}
	\label{fig:TetDiscreteVsContinuum}
\end{minipage}

\end{figure}

Following the projection approach described in Section \ref{subsec:continuoustodiscrete}, we construct the spatially-variant tetragonal lattice from the design variable fields in \autoref{fig:SquareContinuumFields} with different length scales $\mu$. Examples are presented in \autoref{fig:TetragonalBeamMesh}. 

To evaluate the performance of the projection method, we calculate the compliance of the discrete lattice (cf.~\eqref{eq:ComplianceDisc}) subjected to the boundary value problem depicted in \autoref{fig:discreteCantileverBeam}. In \autoref{fig:TetDiscreteVsContinuum} we compare the discrete compliance $\calC_D$ to the previously obtained continuum compliance $\calC_C$ at different regularization intensities. Note that at a length scale of $\mu=1/320$ the lattice contains more than 200{,}000 elements. Direct optimization of such high-resolution discrete structures would be very time-demanding, while the desired resolution follows (almost) for free from this multiscale approach (especially since changing the length scale $\mu$ does not require solving a new topology optimization problem but is merely a change in postprocessing).

The discrete compliance converges towards the continuum approximation for all cases. Depending on the regularization strength, the projected lattice compliance is within $30\%$ of the optimization result for a coarse length scale of $\mu \approx 1/80$ at $c_p=10^{-10}$, $\mu \approx 100$ at $c_p=10^{-11}$, and within $20\%$ of the optimization result for a length scale of $\mu \approx 1/150$ at $c_p=10^{-10}$, and $\mu \approx 1/170$ at $c_p=10^{-11}$. For the lowest regularization ($c_p=10^{-13}$), the discrete compliance is still $>7\%$ higher for the highest considered resolution ($\mu = 1/320$).

The convergence behavior indicates that -- for arbitrary regularization -- the discrete compliance will ultimately near the continuum compliance when the length scale approaches zero ($\mu \to 0$); a small offset is expected to stem from the inevitable differences in both clamping and loading conditions between the formulations of the continuum and the discrete problem (see \autoref{fig:continuumCantileverBeam} and \autoref{fig:discreteCantileverBeam}). 

In summary, optimziation of the tetragonal-lattice cantilever beam reached minimized compliance values close to those of comparable studies in the literature -- but still above the reported minima. This is attributed to a lack of directional stiffness tunability in the tetragonal RUC definition and our choice of $\bflambda$. Therefore, we proceed to enrich the truss design space by considering the orthorhombic lattice family.

\subsection{Orthorhombic lattice family}
\label{Sec:Orthorhombic}
In addition to tailoring the local unit cell orientation through angle $\alpha$ as for the tetragonal lattices in Section~\ref{Sec:TetragonalContinuum}, the orthorhombic lattice family (see \autoref{tab:RUCconstraints}) allows us to tune the anisotropy of the lattice through design variable $\gamma$, as shown in \autoref{fig:ElasticSurfaces}. As before, we first discuss the continuum solution, before demonstrating discrete lattices.

\subsubsection*{Continuum}
\label{Sec:OrthorhombicContinuum}

\begin{figure}
	\centering
\begin{subfigure}[t]{0.26\linewidth}
    	\centering
     	\includegraphics[width=\linewidth]{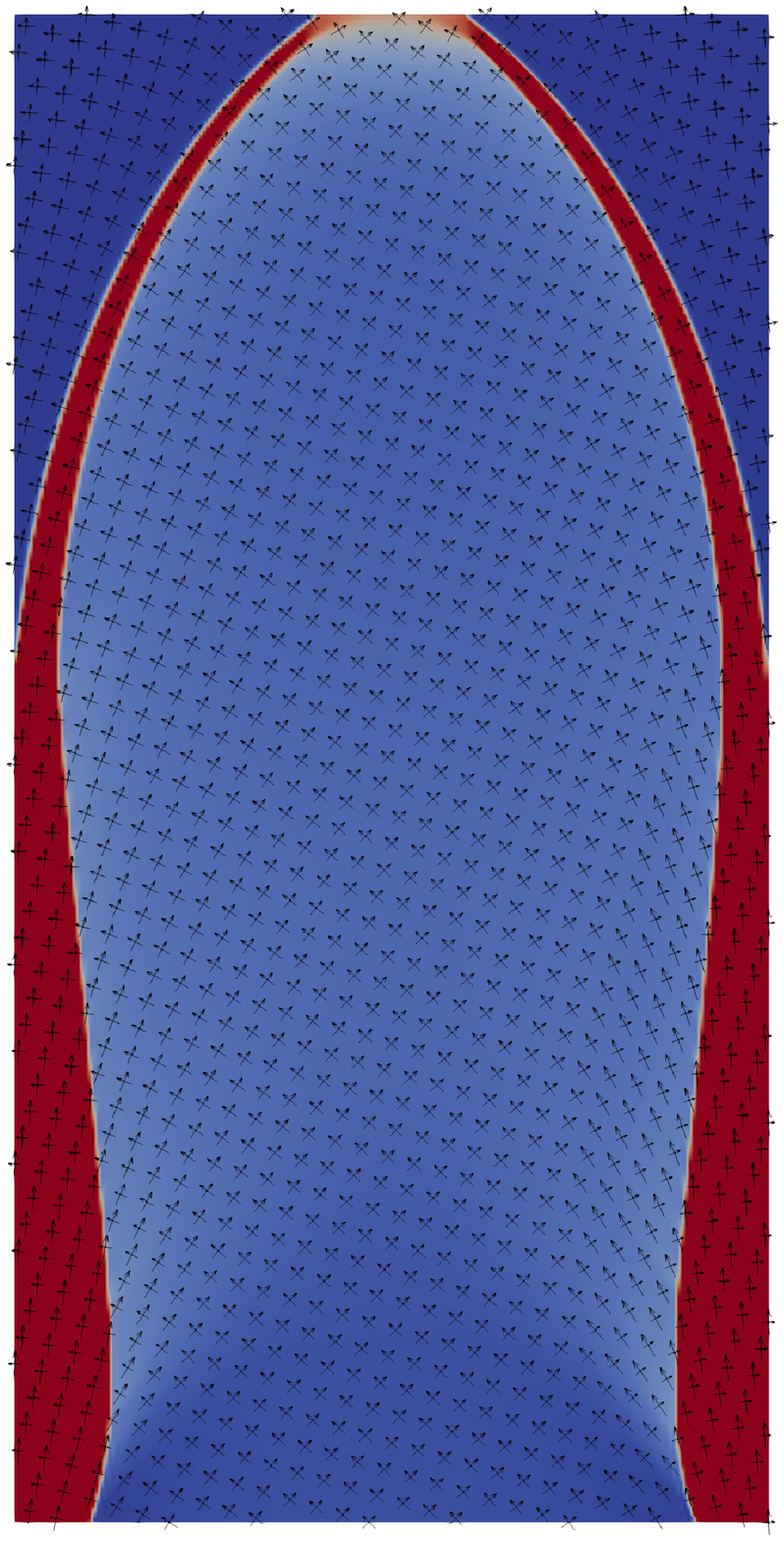}
    	\caption{$c_p=10^{-13}$}
    	\label{fig:Orthorhombic400x200_13}
	\end{subfigure}
	\hfill
	\begin{subfigure}[t]{0.26\linewidth}
    	\centering
     	\includegraphics[width=\linewidth]{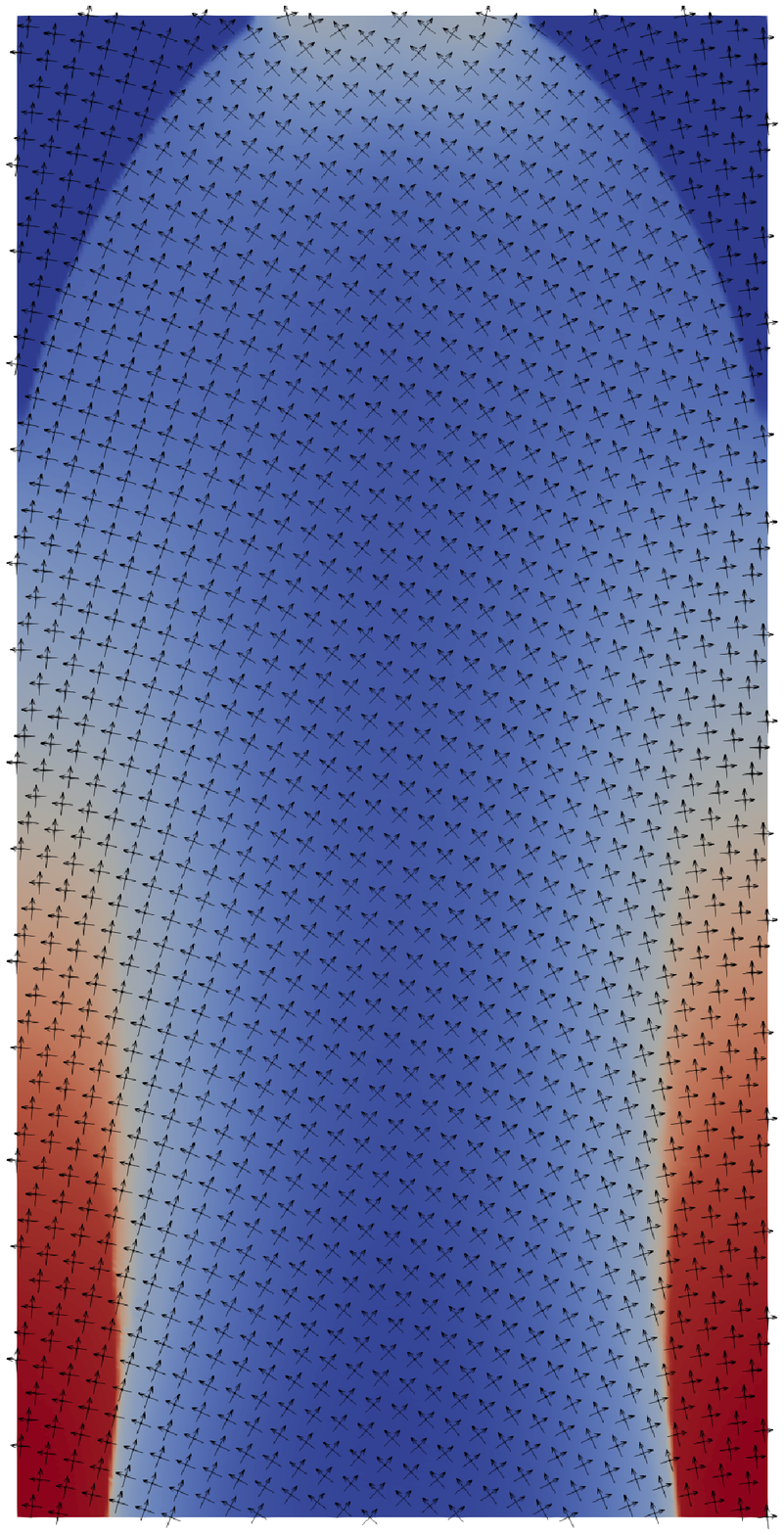}
    	\caption{$c_p=10^{-12}$}
    	\label{fig:Orthorhombic400x200_12}
	\end{subfigure}
	\hfill
	\begin{subfigure}[t]{0.26\linewidth}
    	\centering
     	\includegraphics[width=\linewidth]{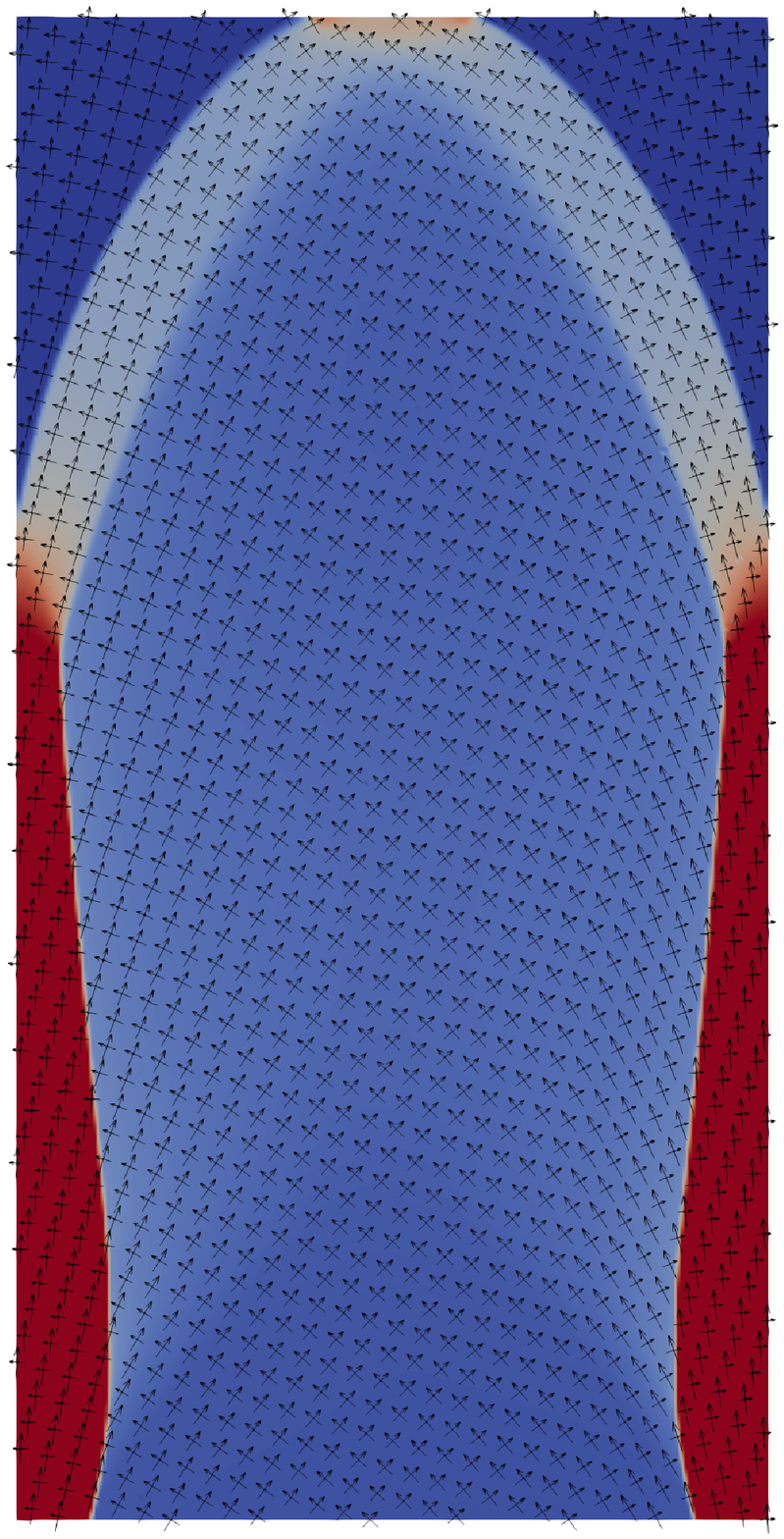}
    	\caption{$c_p=10^{-11}$}
    	\label{fig:Orthorhombic400x200_11}
	\end{subfigure}
	\hfill
	\begin{subfigure}[t]{0.05\linewidth}
    	\centering
     	\includegraphics{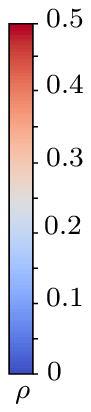}
	\end{subfigure}
	\caption{Optimized orthorhombic-lattice cantilever beam for different regularization constants. Local Bravais bases are indicated by arrows, and the density is shown by the color code. The arrows are scaled and shown at selected locations for readability. Resolution: 400x200 elements. All plots are rotated by 90$^\circ$. Full design variable fields can be found in \autoref{fig:Orthorhombic400x200}.}
	\label{fig:Orthorhombic400x200_2}
\end{figure}

Analogous to Section \ref{Sec:TetragonalContinuum}, we solve the optimization problem \eqref{eq:toproblem}, now subjected to orthorhombic RUC constraints (see \autoref{tab:RUCconstraints}), using a continuum mesh of 400x200 bilinear elements. The resulting design variables for regularization constants $c_p=\{10^{-11},10^{-12},10^{-13}\}$ are presented in \autoref{fig:Orthorhombic400x200}. Examples are shown in \autoref{fig:Orthorhombic400x200_2}. The corresponding compliance values are included in \autoref{tab:MinimizedContinuumCompliance}.˙

In comparison to the tetragonal lattice, the compliance is up to  $7\%$ lower. Comparing to \cite{groen_homogenization-based_2018} (where no local volume constraint was used), the result at $c_p=10^{-13}$ is within $6\%$ difference, which is notable given the constraints on local volume fraction ($\bar\rho(\bfX) \leq 0.5$) shown in \autoref{tab:RUCconstraints}.

These results are promising, but they should be interpreted with caution, as some of the beams in the design become relatively thick (as a consequence of high local densities). This stretches the beam slenderness assumption. We refer to \ref{sec:beam_slenderness} for a detailed study of suitable beam slenderness ratios in compliance with the chosen Timoshenko-Ehrenfest beam model. If we allow for relatively high local densities, then (as shown in \ref{subsec:node_influence}) we accept local beam stiffness overpredictions of around $10\%$. This may result in potentially underestimated structural compliance values. As we are primarily presenting an optimization framework, where homogenization and RUC definition is interchangeable, respectively, we here accept these uncertainties.

\subsubsection*{Discrete lattice}
\label{Sec:OrthorhombicDiscrete}

\begin{figure}
\centering
\begin{minipage}{0.48\textwidth}	
  \centering
     	\includegraphics[width=\linewidth]{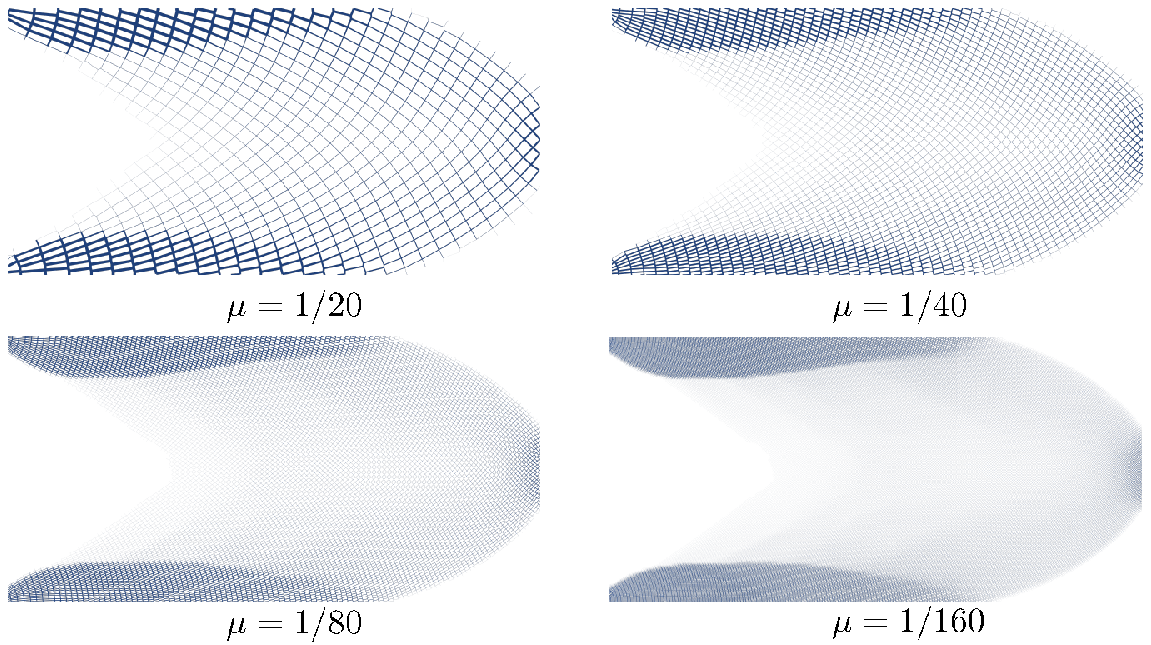}
    \caption{Orthorhombic-lattice cantilever beam at different length scales $\mu$ for $c_p=10^{-12}$.}
    \label{fig:OrthorhombicBeamMesh}    
\end{minipage}%
\hfill
\begin{minipage}{0.48\textwidth}	
  \centering
     	\includegraphics[width=\linewidth]{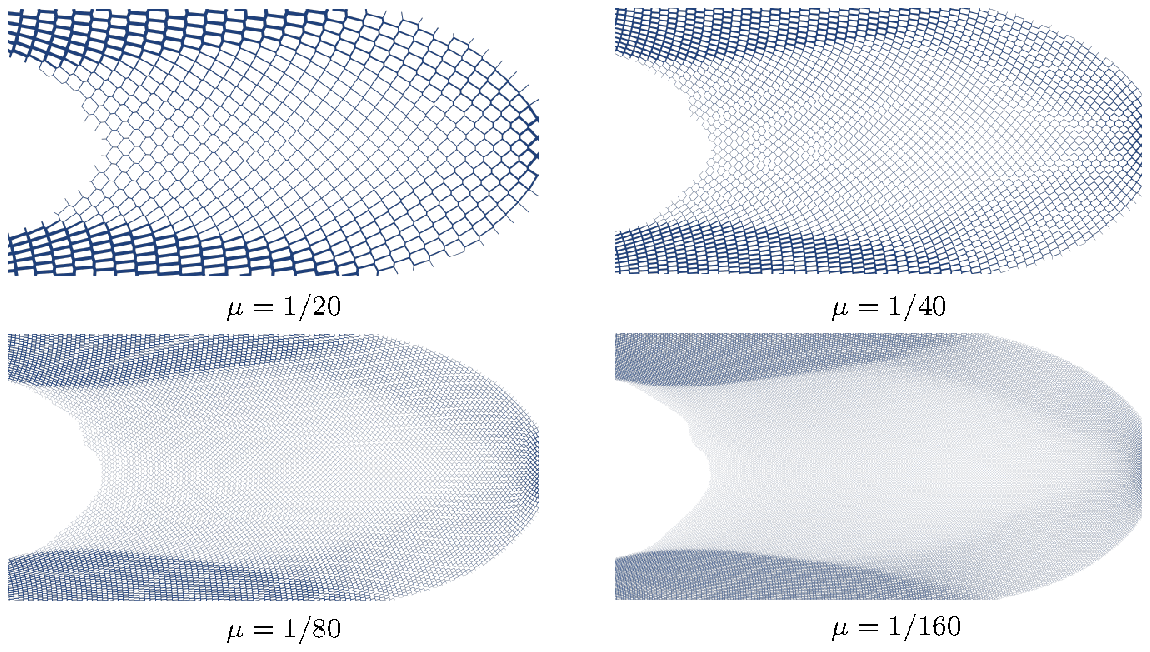}
    \caption{Monoclinic-lattice cantilever beam at different length scales $\mu$ for $c_p=10^{-12}$.}
    \label{fig:MonoclinicBeamMesh}
\end{minipage}

\end{figure}

Analogous to the tetragonal lattice, we project the design variable fields presented in \autoref{fig:Orthorhombic400x200} onto different finite scales. Results are shown in  \autoref{fig:OrthorhombicBeamMesh}. As the $\alpha$-fields in \autoref{fig:Orthorhombic400x200_2} indicated, the projections are similar in orientation to the tetrageonal case. The anisotropy of the orthorhombic lattice compared to the tetragonal lattice is more pronounced but equal in directionality (see \autoref{fig:ElasticSurfaces}). In \cite{pedersen_optimal_1989} it was showon that the optimal orientation of an orthotropic composite coincides with the principal stress directions. Therefore for similar orthotropic directionality the orientation must be same.

In addition, the free design variable $\gamma$ allows to pronounce the directionality which is strongly utilized towards the supports, where the principle stresses are high. Results indicate that a larger range for $\gamma$, allowing for even more strongly pronounced directionality, would allow for further minimization of the compliance. However, a RUC that spans more space in one direction than in the other also requires smaller length scales to approach the continuum compliance. Furthermore, the approximation concerns regarding the chosen homogenization approach based on slender beams (discussed in  \ref{sec:beam_slenderness}) increase with decreasing slenderness. As a trade-off we hence choose to restrict $\gamma$ to the given interval ($0.5 \leq \gamma \leq 2$).

\subsection{Monoclinic lattice family}

In an ultimate step, we enrich the set of design variables $\boldsymbol{\psi}$ by additionally allowing the parameter $\beta$ to be optimized (see \autoref{tab:RUCconstraints}) and study its influence on the truss design and on the effective compliance. $\beta$ not only pronounces the orthotropy (in addition to $\gamma$), but it also changes the topology of the elastic surface and introduces asymmetry (see \autoref{fig:ElasticSurfaces}).

\subsubsection*{Continuum}
\label{Sec:MonoclinicContinuum}

\begin{figure}
	\centering
\begin{subfigure}[t]{0.45\linewidth}
    	\centering
     	\includegraphics[width=\linewidth]{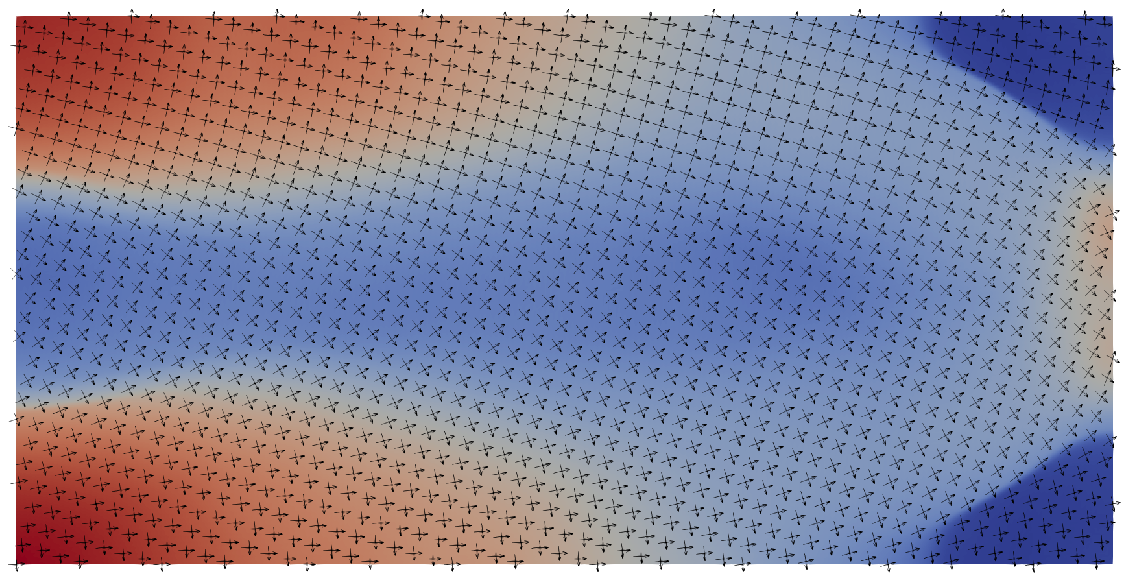}
    	\caption{$c_p=10^{-11}$}
    	\label{fig:Monoclinic400x200_11}
	\end{subfigure}
	\hfill
	\begin{subfigure}[t]{0.45\linewidth}
    	\centering
     	\includegraphics[width=\linewidth]{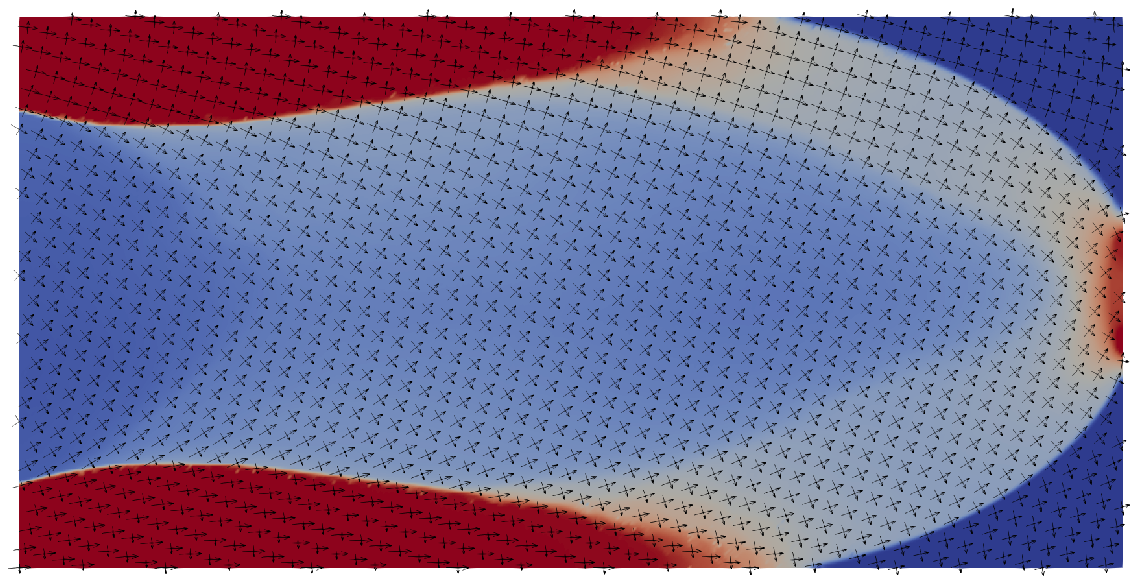}
    	\caption{$c_p=10^{-12}$}
    	\label{fig:Monoclinic400x200_12}
	\end{subfigure}
	\hfill
	\begin{subfigure}[t]{0.03\linewidth}
    	\centering
     	\includegraphics[width=2\linewidth]{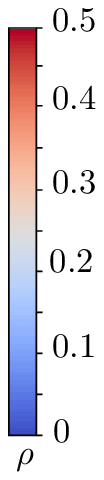}
	\end{subfigure}
	\caption{Monoclinic-lattice cantilever beam for different regularization values. Local Bravais bases are indicated by arrows, and the density is shown by the  color code. The arrows are scaled and shown at selected locations for readability. Resolution: 400x200 elements. Full design variable fields can be found in \autoref{fig:Monoclinic400x200}.}
	\label{fig:Monoclinic400x200_2}
\end{figure}

Analogous to the previous lattice families, we solve the optimization problem \eqref{eq:toproblem}, subjected to the monoclinic RUC constraints of \autoref{tab:RUCconstraints}, on a continuum mesh of 400x200 bilinear elements. The optimized design variable fields for moderate regularization constants $c_p=\{10^{-11},10^{-12}\}$ are summarized in \autoref{fig:Monoclinic400x200} and \autoref{fig:Monoclinic200x100}. Examples are shown in \autoref{fig:Monoclinic400x200_2}. Both configurations emerge from the orthorhombic initial guess. The resulting compliance values are, again, included in \autoref{tab:MinimizedContinuumCompliance}.

In terms of the compliance, the optimized monoclinic lattice performs worse than the optimized orthorhombic and tetragonal lattices when starting from homogeneous initial conditions. By contrast, when using the results from the orthorhombic lattice optimizations as an initial guess, all design variables, including the monoclinic design parameter $\beta$, remain almost unchanged and hence the same compliance is attained. The additional design parameter hence worsens the challenge of finding a good initial guess, since more local minima of the optimization problem arise and the gradient-based optimization approach converges into the closest to the initial guess. Therefore, no solution with better performance than the orthorhombic lattice family could be found. In fact, some of the local minima can be higher than those of the more constrained orthorhombic lattice (cf.\  \autoref{tab:MinimizedContinuumCompliance}). 
This finding is in line with previous studies (e.g. \cite{groen_homogenization-based_2019} and \cite{sigmund_non-optimality_2016}), which  suggested that the orthorhombic lattice family is performing close to theoretical limits. The monoclinic lattice family includes these results and, in principle, better informed initial guesses may lead to an improvement in terms of the global compliance lower than the presented one.

\subsubsection*{Discrete}
\label{Sec:MonoclinicDiscrete}

The projections of the design variable fields in \autoref{fig:Monoclinic400x200} onto different finite scales are shown in \autoref{fig:MonoclinicBeamMesh}. The resulting lattice has a slight asymmetry, due to the asymmetric effective properties of monoclinic RUCs (see \autoref{fig:ElasticSurfaces}) and shows different underlying RUC topologies.

The additional design parameter $\beta$ introduces asymmetry in the RUC anisotropy. This anisotropy reflects in non-rectangular RUCs in the discrete projection and different orientations of the RUCs in comparison to the tetragonal and orthorhombic lattices. The monoclinic lattice family hence shows the most complex lattice architecture yields higher compliance values than previous results.

\section{Conclusion}
\label{sec:conclusion}
Cellular materials show enormous promise as light-weight materials spanning a wide mechanical property space. While the large design freedom allows for target-specific structural integration of spatially variant microstructures, it also poses a challenging inverse-design problem with respect to computational costs and manufacturability.

We presented a two-scale topology optimization scheme for complex macroscale boundary value problems, which guarantees compatible and manufacturable networks of spatially-variant cellular networks. The approach leverages crystallographic principles for continuous and generic unit cell definitions in arbitrary dimensions (here illustrated, without loss of generality, in 2D). The approach is suitable for a wide range of microscale constitutive descriptions, including, e.g., analytical, numerical, or ML-based constitutive laws as well as tabulated material data. We here explored an FE$^2$-type numerical homogenization approach, which derives the effective constitutive response from a discrete structural unit cell on the microscale.

We demonstrated the framework by optimizing spatially-variant trusses in 2D for minimum compliance. A semi-analytical homogenized continuum model allowed for computationally efficient structural optimization on a regular laptop, while providing architectures of high resolution. The assumptions intrinsic in the underlying slender-beam theory put constraints on the beam slenderness and hence on the volume density of the RUC. While the Timoshenko-Ehrenfest beam theory has proven to accurately represent beam behavior at low slenderness, the nodal joints between beams introduce errors. As shown in the Appendix, we investigated the expected resulting error and expect maximum errors of ca.\ $10\%$ for local relative density values up to $50\%$. Even with the imposed constraints, the minimized compliance was within $6\%$ of the comparable (unconstrained) benchmark in \cite{groen_homogenization-based_2018}.

In general, we observed a competition between the regularization of the continuous design variable fields and the length scale of the finite-scale discrete truss in line with the assumption of a separation of scales. For low regularization values, where usually best optimization results are obtained, a very fine length scale is required to ensure similar performance of the discrete lattice versus the continuum prediction. Conversely, if a coarse length scale is required (e.g., by manufacturing constraints), stronger regularization is needed to ensure consistency between the final discrete structure and the continuum approximation.

While the large design freedom in cellular materials offers an enormous mechanical property space, its nonconvex nature requires special attention in topology optimization. Especially, when utilizing a multi-variable parametrization, the optimization is prone to identify only local minima based on the initial guess, which can be far off from the global minimum. To avoid unfavorable local minima, it can be beneficial to explore the design space by optimizing distinct sets of design variables separately before optimizing the whole set. In the case of the monoclinic truss lattice family, we observed that optimization of the (more restricted) orthorhombic lattice family produced better results. We could only reproduce these results with the monoclinic lattice when using the output of the orthorhombic lattice optimization as the initial guess.

Future possible directions include the exploration of different unit cell definitions in 2D and 3D and their use for different target objectives. A natural extension with the utilized nonlinear homogenization framework is the exploration of finite deformations, e.g., for soft robotic applications.

\begin{acknowledgment}
O.S.~was supported by the Villum Foundation through the Villum Investigator Project ``InnoTop''.
\end{acknowledgment}

\bibliographystyle{asmems4}

\appendix
\section{Homogenization based on beams}
\label{sec:beam_slenderness}
The utilized homogenization approach of \cite{glaesener_continuum_2019} is based on modeling the RUC by beam elements, which imply rigid (welded) connections at junctions. Beam theory is generally deemed applicable whenever the length of a beam is much larger than its width, i.e., when the slenderness $L/w \gg 1$. In addition, assuming rigid connections between beams neglects nodal effects such as stress concentrations at nodes or overlaps between connected beams. These assumptions, however, are challenged when modeling RUCs with high volume fractions $\bar \rho$, implying relatively thick beams. To evaluate the applicability of the applied homogenization approach, we present in this section a discussion of the influence of the effective RUC density on the approximation of homogenized quantities and ways to extend the applicable range. 
For all calculations we assume a rectangular beam cross-section with out-of-plane thickness $t=1$, Young's modulus $E=1$, Poisson's ration $\nu=0.3$, and the Timoshenko shear coefficient for rectangular cross-sections $\kappa=1.2$ (all normalized for simplicity).

\subsection{Beam slenderness}
\label{subsec:beam_slenderness}

As the beam is the smallest building block of the RUC, the exactness of the beam element has primary relevance. Generally well-known beam formulations include the theories of Euler-Bernoulli and Timoshenko-Ehrenfest. To evaluate their accuracy for low slenderness ratios $L/w$, we evaluate, using both formulations, the strain energy of a 2D cantilever beam of length $L=1$ and varying thickness $w$. The beam is clamped at one end and loaded by a constant displacement at the other end. For comparison, we establish a reference by resolving the cantilever beam ($20$ elements over the width $w$) with square bilinear finite elements, assuming isotropic linear elasticity under the plane-stress assumption. 
          
\begin{figure}[!b]
	\centering
	\begin{minipage}[c]{0.48\textwidth}	
     	\includegraphics[width=0.9\linewidth]{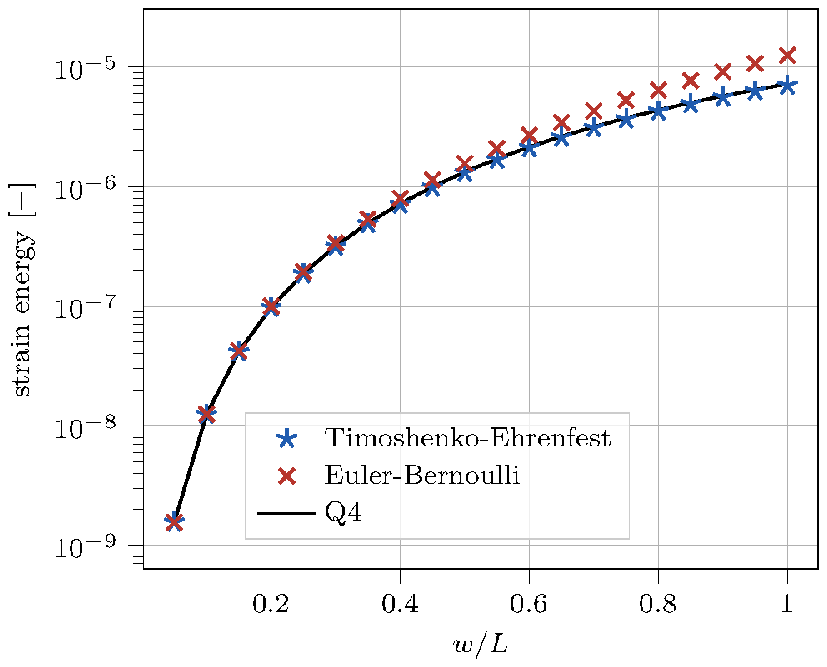}
		\caption{Total strain energy of a cantilever beam (clamped at one end, loaded by a constant deflection at the other end) for varying slenderness $w/L$. The fully resolved result is shown in black (solid line), compared to results obtained from Timoshenko-Ehrenfest beam theory (blue stars) and Euler-Bernoulli beam theory (red crosses).}
		\label{fig:beams_strainEnergy}
	\end{minipage}%
	\hfill
	\begin{minipage}[c]{0.48\textwidth}
		\centering
     	\includegraphics[width=0.8\linewidth]{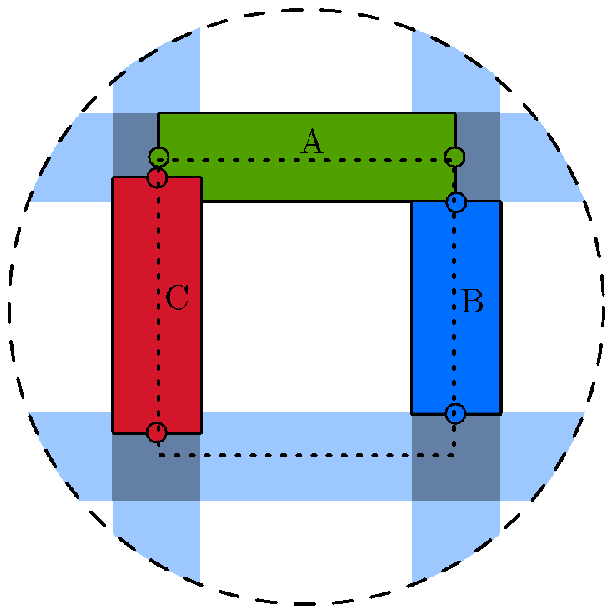}
		\caption{Different node treatments in a square RUC ($\bar\rho=0.5$): variant A ignores the overlap of beams at a node (dark areas) and uses a beam element from node center to node center. In variant B the nodes are assumed rigid and beams have the effective length between nodes without overlap. Variant C is a volume-preserving compromise between variants A and B.}
		\label{fig:node_discussion}
	\end{minipage}
\end{figure}

A comparison of the strain energy inside the bent cantilever beam is shown in \autoref{fig:beams_strainEnergy}. Both beam implementations converge to the exact solution for $L/w \gg 1$. With increasing $w/L$, Euler-Bernoulli beam theory exceedingly overestimates the strain energy. In contrast, Timoshenko-Ehrenfest beam theory shows significantly better agreement and generally tends to underestimate the strain energy. For $w/L=0.5$, Euler-Bernoulli beam theory overstimates the energy by roughly $27\%$, while Timoshenko-Ehrenfest stays within a $5\%$ error margin for the full range of slenderness ratios up to $w=L$.

We conclude that Euler-Bernoulli beam theory is not suited for modeling beams with low slenderness ratio $L/w$. In contrast, when accepting a maximum error of $5\%$, Timoshenko-Ehrenfest beam theory is applicable for a wide range of slenderness ratios and even for extreme cases ($w=L$). Therefore, all following discussions (and the truss optimization studies) are based on Timoshenko-Ehrenfest beam theory.

\subsection{Node influence}
\label{subsec:node_influence}

Truss networks consist of many beams that are connected to each other through nodes. While the previous section confirmed the validity of beam theory for an individual beam, we here investigate the influence of nodes on the approximation of homogenized properties using beam elements. Generally, the influence of nodes is neglected when using beam models, assuming a slenderness ratio of $L/w \gg 1$. However, when beams are not slender, or equivalently when the density $\bar\rho$ of the RUC is high, the influence of the nodes on the effective mechanical properties cannot be neglected. 

In \autoref{fig:node_discussion}, we introduce three variants for approximating the geometry of a RUC with non-slender beams. As in classical beam theory, all variants assume that the node is rigid and hence described well by Timoshenko-Ehrenfest theory. In variant A, beams connect from node centroid to node centroid; beams overlap at nodes (this is the standard approach). In variant B, the overlapping nodes are accounted for by assuming beams whose effective length is not from node centroid to node centroid but the reduced length between nodes without overlapping volumes. Variants A and B present the soft and stiff extremes (lower and upper bounds), respectively. Finally, variant C is a compromise, where the overlap of beams at each node is chosen such that the volume of all beams in a RUC matches the volume fill fraction according to the effective density $\bar \rho$, which defines the effective beam length (between those of variants A and B).  Note that, when $L/w \gg 1$, all three variants must converge to the same solution.

\begin{figure}[!b]
	\centering
	\begin{subfigure}[t]{0.23\linewidth}
	 \centering
    \includegraphics[width=\linewidth]{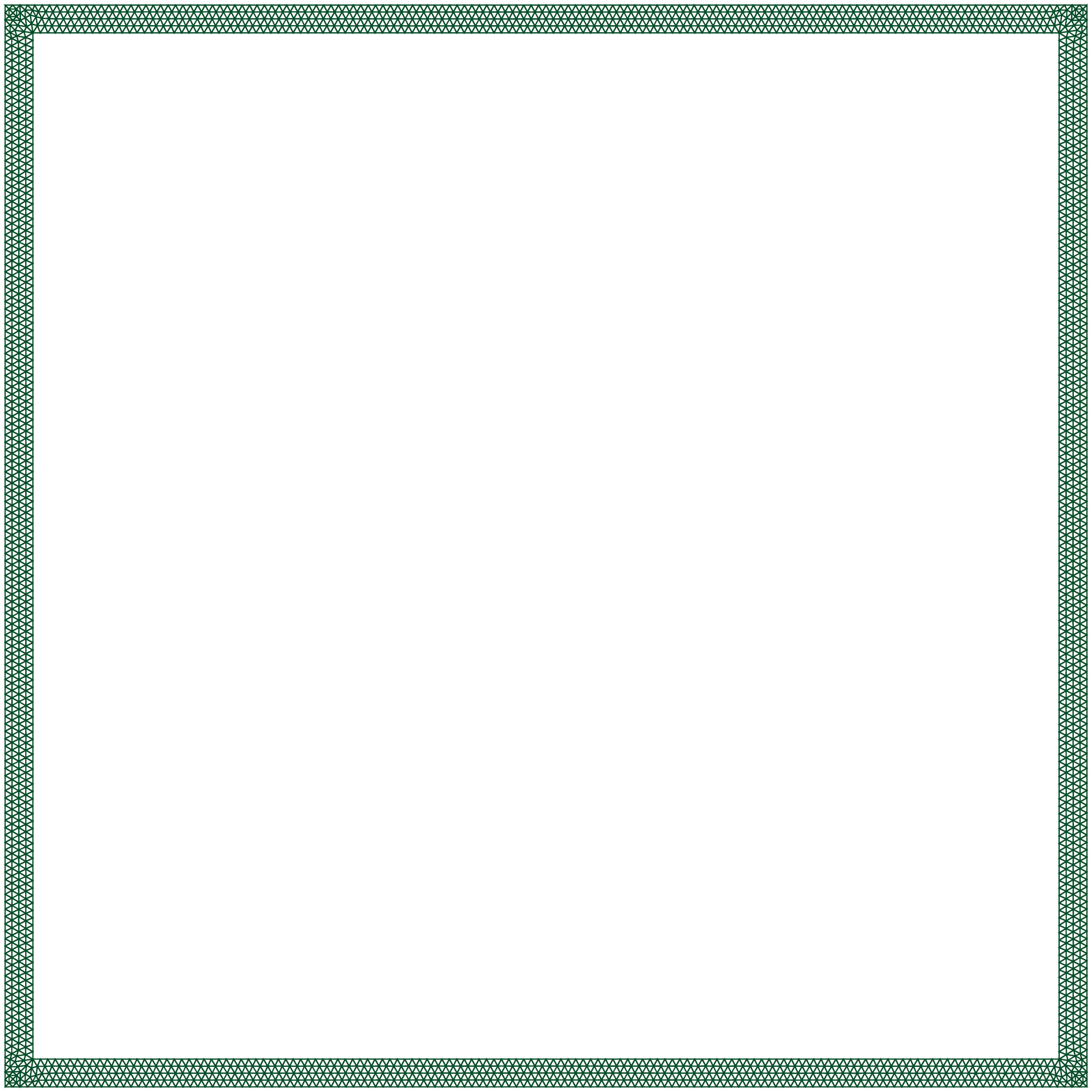}
  \caption{square RUC, $\bar{\rho}=0.1$}
		\label{fig:squareUC0.1}
\end{subfigure}%
	\hfill
	\begin{subfigure}[t]{0.23\linewidth}
	 \centering
    \includegraphics[width=\linewidth]{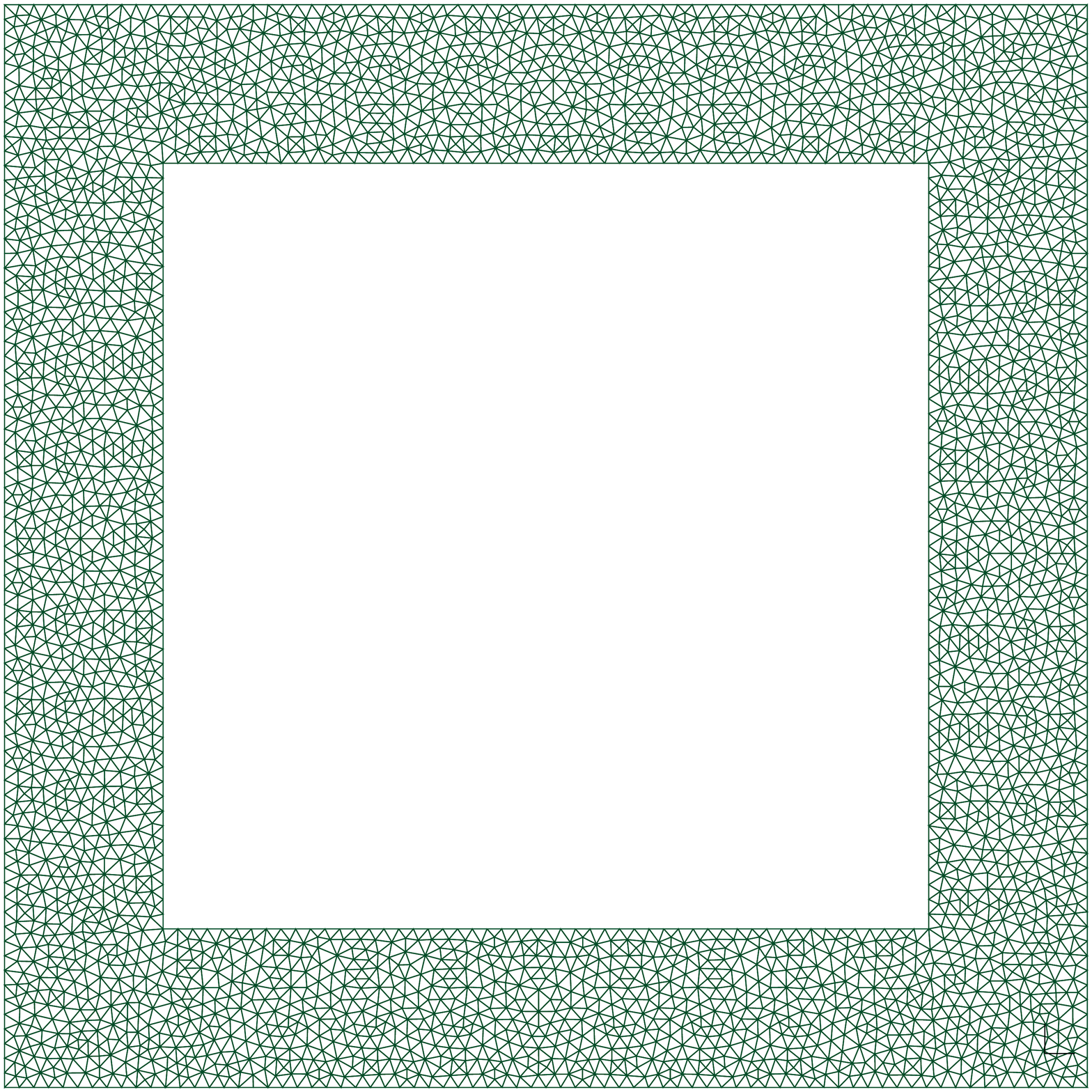}
  \caption{square RUC, $\bar{\rho}=0.5$}
		\label{fig:squareUC0.5}
\end{subfigure}%
	\hfill
	\begin{subfigure}[t]{0.23\linewidth}
	 \centering
    \includegraphics[width=\linewidth]{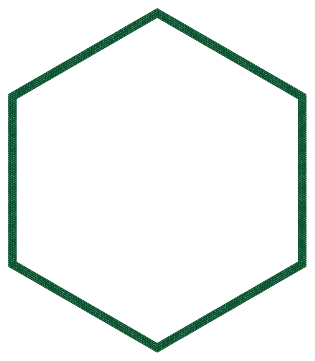}
  \caption{hexagon RUC, $\bar{\rho}=0.1$}
		\label{fig:hexUC0.1}
\end{subfigure}%
	\hfill
	\begin{subfigure}[t]{0.23\linewidth}
	 \centering
    \includegraphics[width=\linewidth]{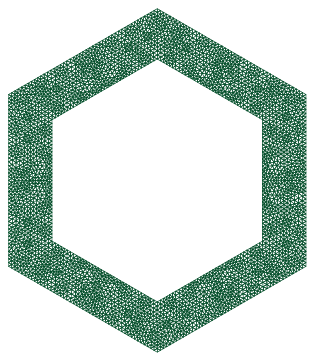}
  \caption{hexagon RUC, $\bar{\rho}=0.5$}
		\label{fig:hexUC0.5}
\end{subfigure}

	\begin{subfigure}[t]{0.23\linewidth}
    \includegraphics[width=\linewidth]{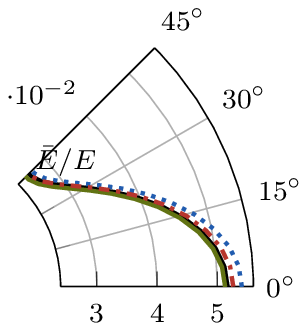}
    \caption{square RUC, $\bar{\rho}=0.1$}
	\label{fig:elastic_surface_square_0.100000}
	\end{subfigure}%
	\hfill
	\begin{subfigure}[t]{0.23\linewidth}
	
    \includegraphics[width=\linewidth]{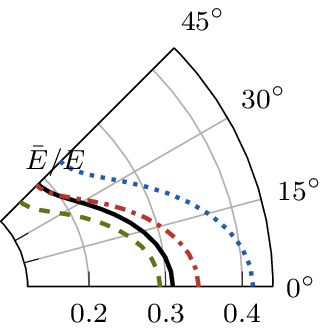}
  \caption{square RUC, $\bar{\rho}=0.5$}
		\label{fig:elastic_surface_square_0.500000}
	\end{subfigure}%
	\hfill
	\begin{subfigure}[t]{0.23\linewidth}
	
		\centering
    \includegraphics[width=\linewidth]{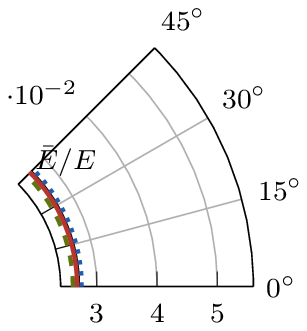}
  \caption{hexagonal RUC, $\bar{\rho}=0.1$}
		\label{fig:elastic_surface_hex_0.100000}
	\end{subfigure}%
	\hfill
	\begin{subfigure}[t]{0.23\linewidth}
	\centering
    \includegraphics[width=\linewidth]{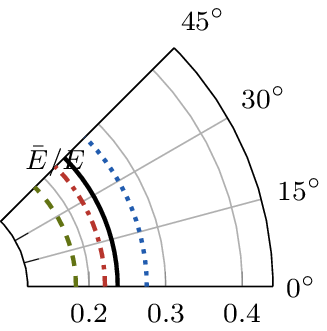}
    \caption{hexagonal RUC, $\bar{\rho}=0.5$}
		\label{fig:elastic_surface_hex_0.500000}
\end{subfigure}%
\caption{Influence of the node treatment on the effective directional Young's modulus for different effective RUC densities and topologies. Figures \ref{fig:squareUC0.1} to \ref{fig:hexUC0.5} show the CST mesh of the square RUC ($\alpha=0$, $\beta=\pi/2$, $\gamma=1$) and hexagonal RUC ($\alpha=0$, $\beta=\pi/3$, $\gamma=1$), each for $\bar\rho=0.1$ and $\bar\rho=0.5$. Figures \ref{fig:elastic_surface_square_0.100000} to \ref{fig:elastic_surface_hex_0.500000} show the respective directional Young's moduli based on homogenization of the shown RUCs (black solid lines) and obtained from the modified beam representation in variant A, ignoring the overlap at nodes (green dashed lines), assuming rigid nodes in variant B (blue dotted lines), and volume-preserving node overlaps in variant C (red dash-dotted lines). Expoiting symmetry, only half of the first quadrant is shown in each elastic modulus plot.}
\label{fig:nodeStiffness}
\end{figure}

The influence of the three variants on the effective stiffness is demonstrated for the square and hexagon RUCs, as representatives of stretching- and bending-dominated RUCs, in \autoref{fig:nodeStiffness}. As a reference, we calculated the effective directional Young's modulus of fully discretized (ref) RUCs, also shown in Figures \ref{fig:squareUC0.1} to \ref{fig:hexUC0.5}, with classical homogenization assuming periodic boundary conditions and linear elasticity under the plane-stress assumption. These reference values are indicated as black solid lines. The directional Young's modulus based on truss homogenization \citep{glaesener_continuum_2019} for the three variants A, B, C are shown in green, blue, and red, respectively. 

The results confirm that for low effective density (e.g., $\bar\rho=0.1$) all three variants converge to the fully-resolved reference solution. Variant A converges from below, variant B converges from above, and there is no clear trend for variant C. For the large density $\bar\rho=0.5$, the trends are the same, but the deviations are more severe. Variant A underestimates the directional stiffness of the square RUC by maximally $5\%$, variants B and C overestimate the directional stiffness by up to $34\%$ and $11\%$, respectively. For the hexagonal RUC, variant B still overestimates the directional stiffness by $16\%$, and variants A and C underestimate the stiffness by $23\%$ and $7\%$, respectively. 

\begin{figure}
	\centering
	\begin{subfigure}[t]{0.33\linewidth}
		\centering
        \includegraphics[width=\linewidth]{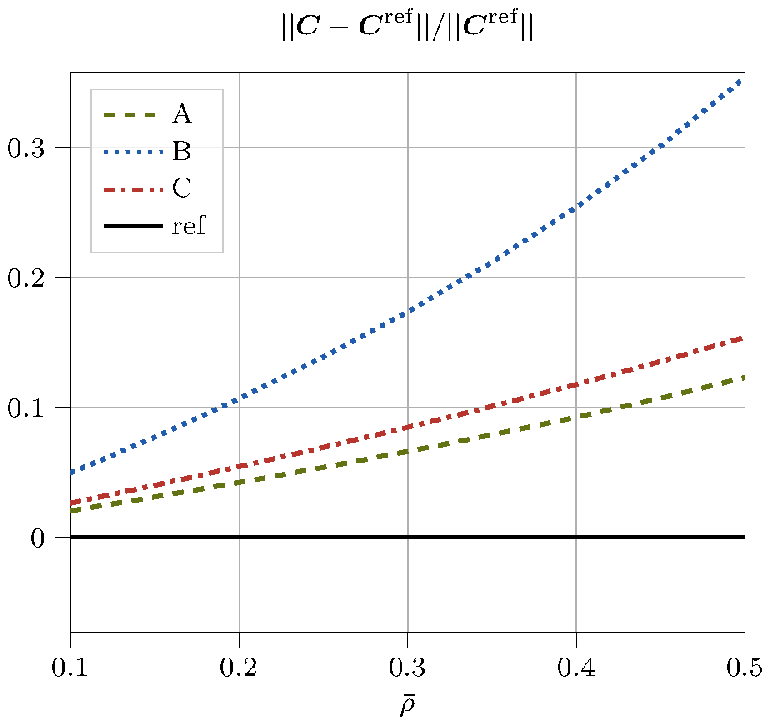}
		\caption{square RUC}
		\label{fig:Cnorm_squareUC}
	\end{subfigure}%
	\hfill
	\begin{subfigure}[t]{0.33\linewidth}
		\centering
        \includegraphics[width=\linewidth]{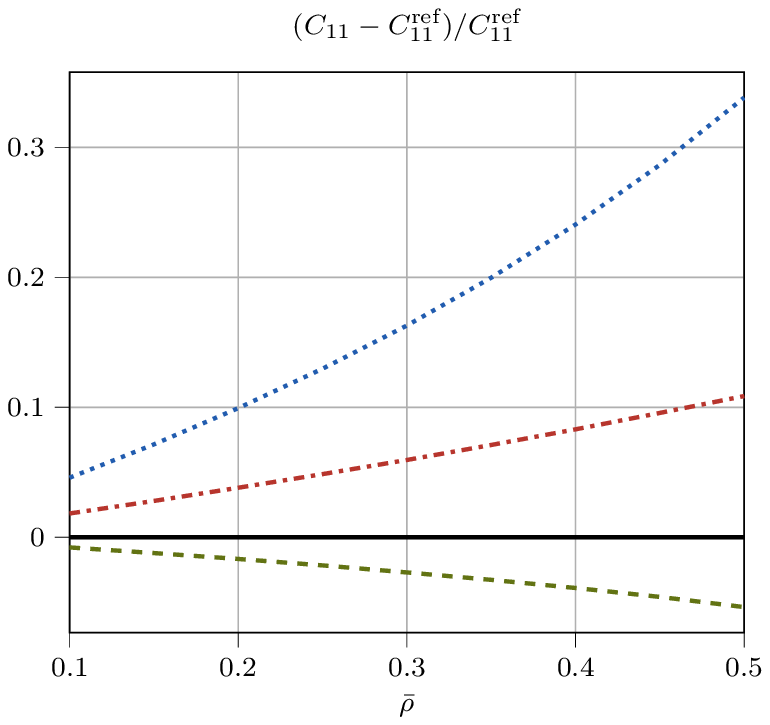}
		\caption{square RUC}
		\label{fig:C00_squareUC}
	\end{subfigure}%
	\hfill
	\begin{subfigure}[t]{0.33\linewidth}
		\centering
        \includegraphics[width=\linewidth]{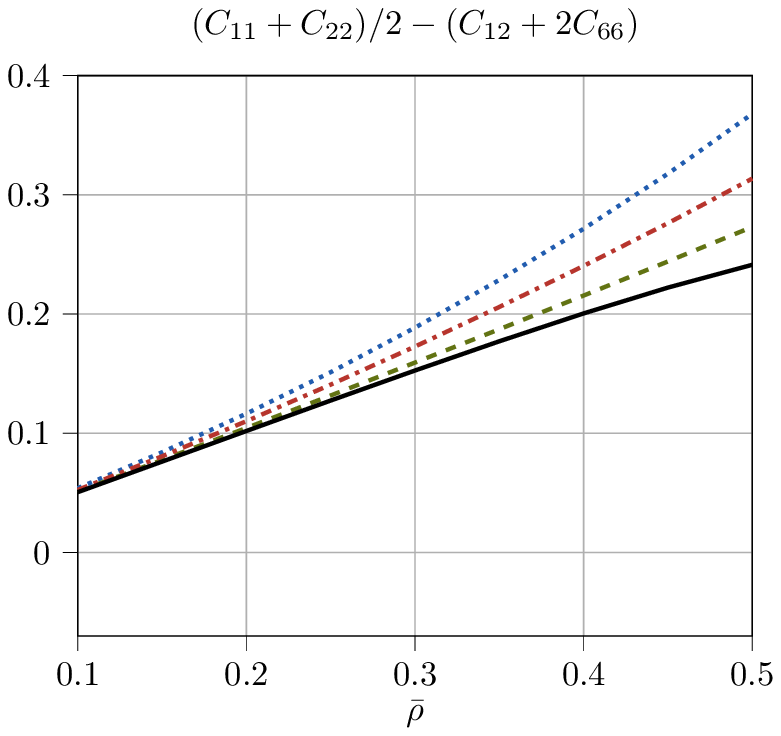}
		\caption{square RUC}
		\label{fig:shearStiffness_sq}
	\end{subfigure}%

	\begin{subfigure}[t]{0.33\linewidth}
		\centering
        \includegraphics[width=\linewidth]{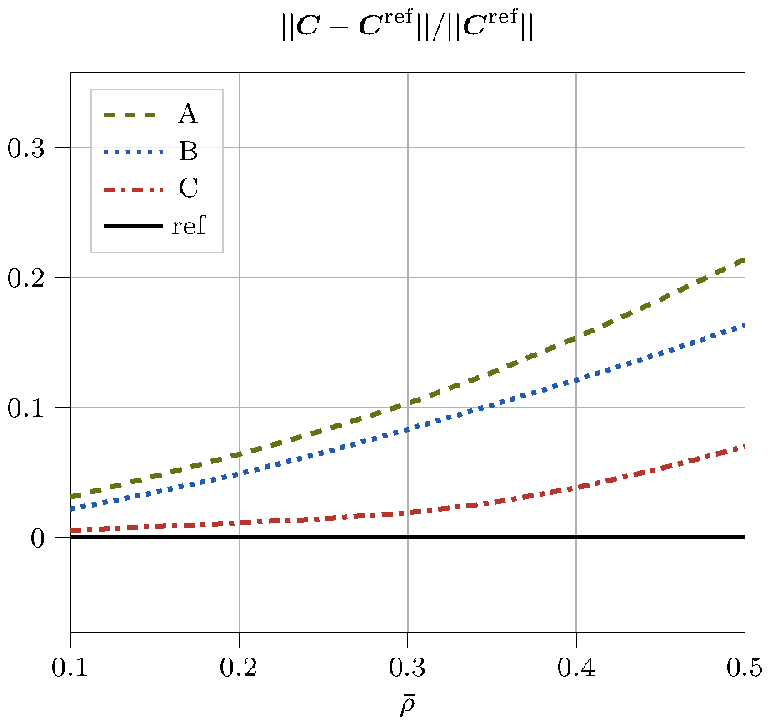}
		\caption{hexagon RUC}
		\label{fig:Cnorm_hexUC}
	\end{subfigure}%
	\hfill
	\begin{subfigure}[t]{0.33\linewidth}
		\centering
        \includegraphics[width=\linewidth]{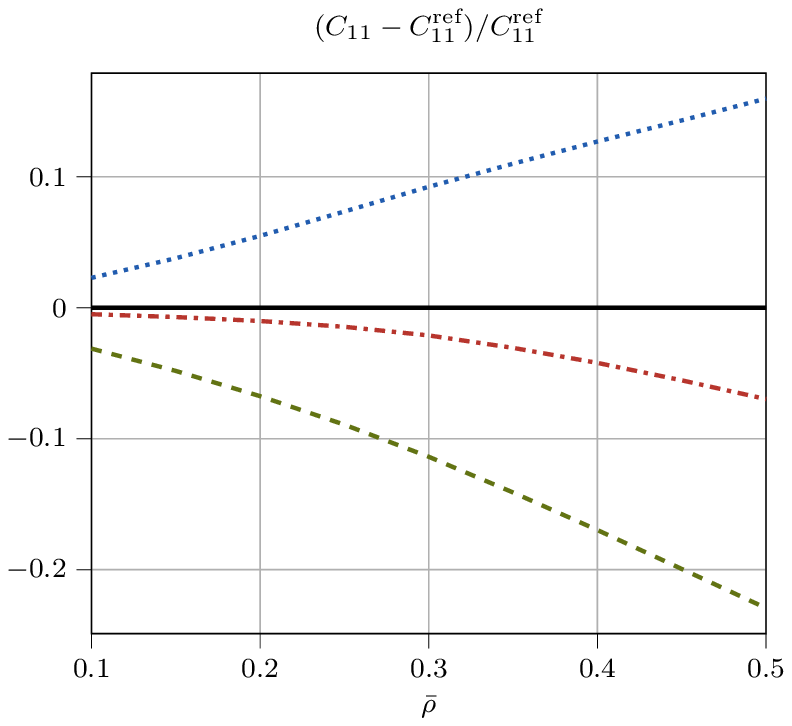}
		\caption{hexagon RUC}
		\label{fig:C00_hexUC}
	\end{subfigure}%
	\hfill
	\begin{subfigure}[t]{0.33\linewidth}
		\centering
		\centering
        \includegraphics[width=\linewidth]{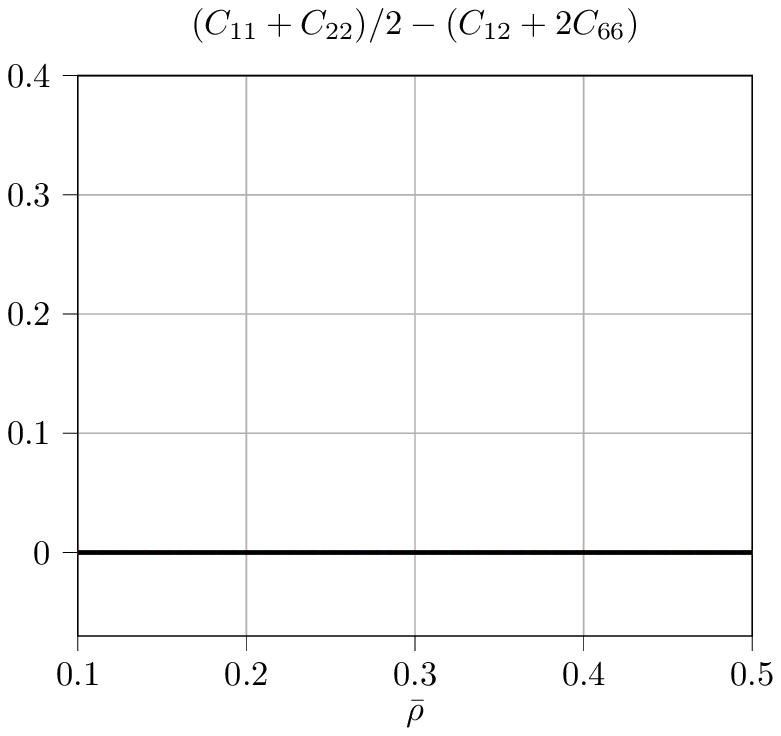}
		\caption{hexagon RUC}
		\label{fig:shearStiffness_hex}
	\end{subfigure}%
	\caption{Influence of effective density and node treatment on estimating the elasticity tensor, comparing the results of beam elements of variants A, B, and C to the fully-meshed reference solution (using Voigt notation in 2D for the stiffness tensor components).}
	\label{fig:stiffnessRUCs}
\end{figure}

The comparison is extended to the full homogenized orthotropic elasticity tensor $\bfC$ of the RUC (having $3\times3$ components, using Voigt notation in 2D) . Results are shown in \autoref{fig:stiffnessRUCs} with the fully discretized reference solution indicated by $C^\text{ref}$. As expected, the errors shown in \autoref{fig:Cnorm_squareUC} and \autoref{fig:Cnorm_hexUC} converge to zero in the limit of low densities $\bar\rho$. For $\bar\rho=0.5$ the error of variant B ($>30\%$) is highest in the square, and variant A shows the largest error in the hexagon ($>20\%$). For the $C_{11}$-entry, \autoref{fig:C00_squareUC} and \autoref{fig:C00_hexUC}) confirm the observations from the directional Young's moduli (cf.~\autoref{fig:nodeStiffness}): with decreasing $\bar\rho$, variant A converges from below, variant B from above, and variant C is scattered around the reference values. Lastly, we compare a measure for the shear stiffness in form of invariant $(C_{11}+C_{22})/2-(C_{12}+2C_{66})$, as presented by \cite{pedersen_optimal_1989}. When the invariant assumes a negative value, the shear stiffness is regarded as relatively high and could indicate artificial shear stiffening. By contrast, a positive value indicates a relatively low shear stiffness. In  \autoref{fig:shearStiffness_hex}, we observe a constant zero invariant for the hexagon, whereas the square RUC loses relative shear stiffness with increasing $\bar\rho$, as shown in \autoref{fig:shearStiffness_sq}.

\subsection{Applicability of beam-based homogenization}

We showed in \ref{subsec:beam_slenderness} that the Timoshenko-Ehrenfest beam theory provides accurate results up to low slenderness values (thick beams), so that we deem this model applicable for the homogenization of RUCs with relatively high filling fraction $\bar\rho$. In \ref{subsec:node_influence}, we observed that the nodes in a RUC can be assumed rigid for relatively low $\bar\rho$. However, the error increases drastically for increasing $\bar\rho$. Clear trends are observable: Variant A consistently underestimates the reference, while variant B consistently overestimates the effective properties. Variant C lies in between those two bounds. No artificial effects, such as shear stiffening, could be detected.

In conclusion, one must carefully evaluate the acceptable error when choosing to homogenize RUCs with high $\bar\rho$ while assuming rigid nodes (as done in this study based on Timoshenko-Ehrenfest beam theory). For this optimization study, we chose to use variant C as a compromise and accept maximum local errors on the order of $~10\%$ by allowing for local relative densities $\rho(\bfX)$ up to $50\%$.

\section{Continuum design variable fields}
\label{sec:continuum_fields}
\begin{figure}
	\centering
        \includegraphics[width=\linewidth]{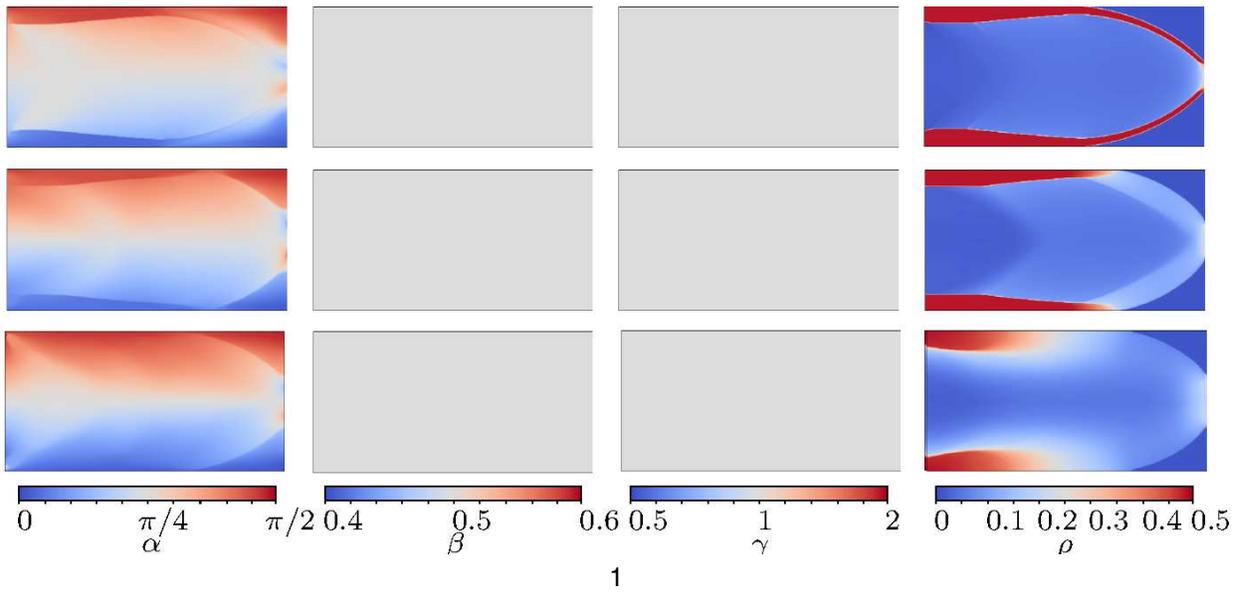}
1	\caption{Optimized design variable fields for the tetragonal-lattice cantilever problem, using 400x200 elements. Top: $c_p=10^{-13}$; middle: $10^{-12}$; bottom: $10^{-11}$.}
		\label{fig:SquareContinuumFields}
\end{figure}

\begin{figure}
    \centering
        \includegraphics[width=\linewidth]{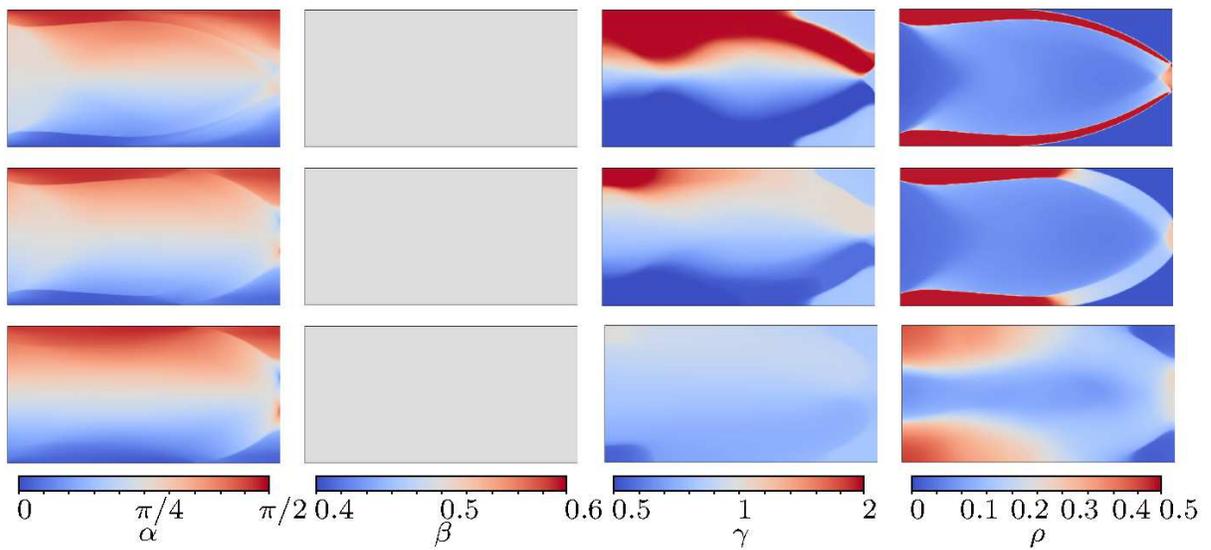}
    \caption{Optimized design variable fields for the orthorhombic-lattice cantilever problem, using 400x200 elements. From top to bottom, $c_p=(10^{-13},10^{-12},10^{-11})$.}
    \label{fig:Orthorhombic400x200}
\end{figure}

\begin{figure}
    \centering
        \includegraphics[width=\linewidth]{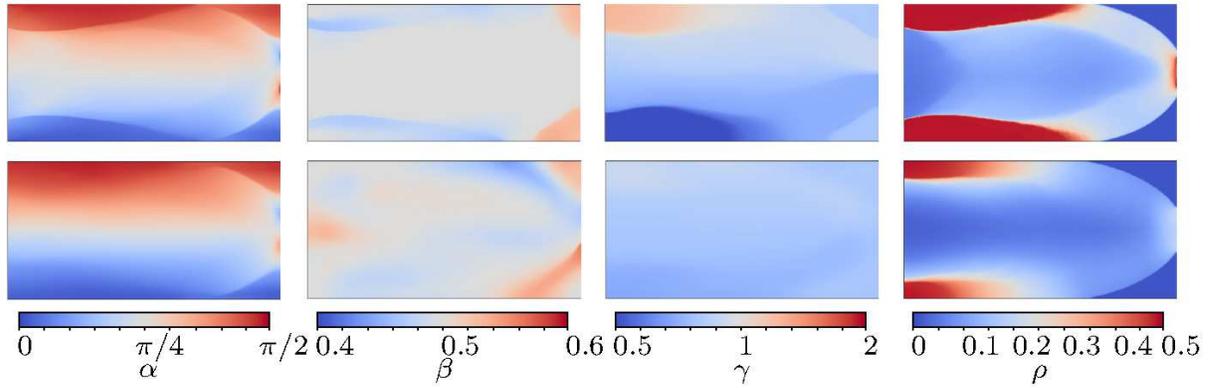}
    \caption{Optimized design variable fields for the monoclinic-lattice cantilever problem, using 400x200 elements and the optimized fields of the for the orthorhombic lattice family as starting guess. From top to bottom, $c_p=(10^{-12},10^{-11})$.}
    \label{fig:Monoclinic400x200}
\end{figure}

All studied design variable fields are shown in Figures \ref{fig:SquareContinuumFields}, \ref{fig:Orthorhombic400x200} and \ref{fig:Monoclinic400x200}.

\begin{figure}
    \centering
        \includegraphics[width=\linewidth]{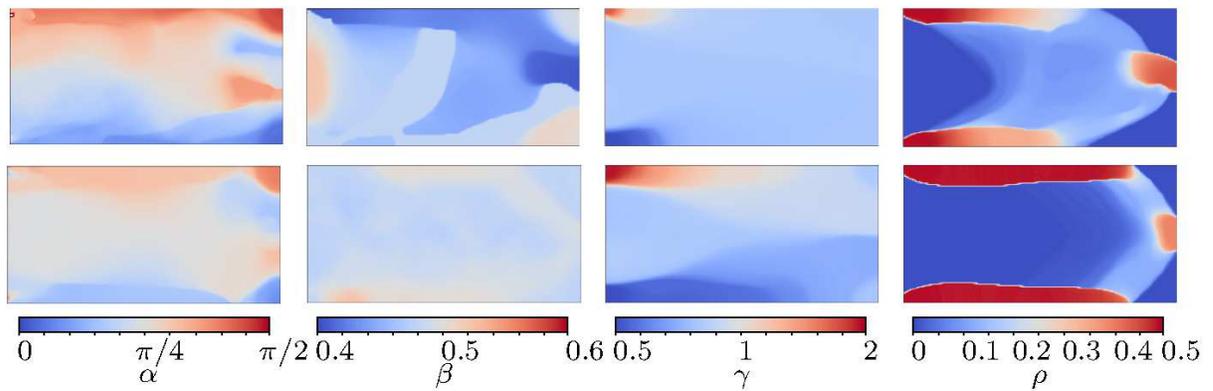}
    \caption{Optimized design variable fields for the monoclinic-lattice cantilever problem, using 400x200 elements and the homogeneous starting guess. From top to bottom, $c_p=(10^{-12},10^{-11})$.}
    \label{fig:Monoclinic200x100}
\end{figure}

\end{document}